\newcommand{\Scalar}{\hat{A}}
\newcommand{\Vector}{\hat{V}}
\newcommand{\Tensor}{\hat{T}}

\newcommand{\HH}{{\cal H}}
\newcommand{\Oop}{{\hat{\cal O}}}
\newcommand{\Hop}{{\hat{\HH}}}
\newcommand{\HMF}{{\Hop_{\rm MF}}}
\newcommand{\EMF}{{E_{\rm MF}}}
\newcommand{\Heff}{{\Hop_{\rm eff}}}
\newcommand{\Heffscalar}{{\HH_{\rm eff}}}
\newcommand{\dH}{{\delta\Hop}}
\newcommand{\dHsame}{{\delta\Hop_{\rm same}}}
\newcommand{\dHdiff}{{\delta\Hop_{\rm diff}}}

\newcommand{\Hone}{{\HH_1(\hat\nn)}}
\newcommand{\HoneQ}{{\Hop_1(\SS)}}

\renewcommand {\ss} {{\bf s}}
\newcommand {\nn} {{\bf n}}
\renewcommand {\SS} {{\bf S}}
\newcommand {\Nsub}{\tilde{N}}
\newcommand {\Ssub}{\tilde{S}}
\newcommand {\tJ} {{\tilde{J}}}
\newcommand {\pp} {{\phi}}
\newcommand {\psii} {{\psi}}
\newcommand {\PP} {{\Phi}}
\newcommand {\nth} {{n{\rm -th}}}
\newcommand {\mth} {{m{\rm -th}}}
\newcommand {\ee} {{\hat{\bf e}}}
\newcommand{\RC}{{\cal R}}

\newcommand {\zab} {{z^m_{\alpha\beta}}}
\newcommand {\Sab} {{\SS_\alpha\cdot\SS_\beta}}

\newcommand {\mm} {{\hat{\bf m}}}

\newcommand {\rr} {{\bf r}}
\newcommand {\QQ} {{\bf Q}}

\newcommand {\BS}{BS~}

\def\REM#1 {{[\bf #1 ]}}
\def\NOTE#1 {{}}

\documentstyle[twocolumn,prb,aps,epsf,eqsecnum]{revtex}

\begin{document}

\twocolumn[\hsize\textwidth\columnwidth\hsize\csname@twocolumnfalse%
\endcsname

\draft

\title{Effective Hamiltonian and low-lying energy clustering patterns
of four-sublattice antiferromagnets}

\author{ N.-G.~Zhang$^a$, 
C.~L.~Henley$^{a}$,
C.~Rischel$^b$, and K.~Lefmann$^c$}
\address{$^a$Dept. of Physics, Cornell University, Ithaca NY 14853-2501}
\address{$^b$ 
Niels Bohr Institute, University of Copenhagen, Blegdamsvej 17, 
DK-2100 København Ø, Denmark
}
\address{$^c$ 
Department of Materials Research,
Ris\o\ National Laboratory, DK-4000 Roskilde, Denmark}
\maketitle

\begin{abstract}
We study the low-lying energy clustering patterns of 
quantum antiferromagnets 
with $p$ sublattices (in particular $p=4$).
We treat each sublattice as a large spin, 
and using second-order degenerate perturbation theory, 
we derive the effective (biquadratic) Hamiltonian coupling
the $p$ large spins.
In order to compare with exact diagonalizations,
the Hamiltonian is explicitly written for a finite-size lattice,
and it contains information on energies of excited states as well as
the ground state.
The result is applied to the face-centered-cubic Type I
antiferromagnet of spin 1/2, including second-neighbor
interactions. A 32-site system is exactly diagonalized, and
the energy spectrum of the low-lying singlets follows the analytically 
predicted clustering pattern.
\end{abstract}
]


\narrowtext  

\section{introduction}

Many frustrated isotropic antiferromagnets 
develop long-range order, but their classical ground states
have nontrivial degeneracies,  so that the
true (quantum) spin ordering pattern is not obvious and may be
decided by small perturbations. 
Exact diagonalization calculations have been carried out
on finite systems and the low-lying eigenenergies show
interesting clustering patterns.
In this paper we develop an effective Hamiltonian approach to
studying energy clustering for a finite-size lattice.
The approximation that we use is to 
divide the system into sublattices following the classical ground states,
represent the spins in each sublattice 
by one large spin, and account for the fluctuations within 
sublattices by an effective Hamiltonian, which couples the
large sublattice spins to each other. 
This effective Hamiltonian is much more manageable
(as far as exact diagonalization is concerned)
than the original Hamiltonian.
From this effective Hamiltonian, 
using notions of Bohr-Sommerfeld quantization, 
tunneling, and the geometrical phase, one can predict the 
energy patterns of low-lying singlet states of a small system, as
may be ``observed'' in exact diagonalizations. 
Recently, in Ref.~\onlinecite{4spin}, two of us studied
the triangular antiferromagnet with nearest and
next nearest neighbor interactions and compared
analytical predictions with exact diagonalization results
in Ref.~\onlinecite{lech95}.
In this paper, we present a systematic derivation of
the (biquadratic) effective Hamiltonian for a class
of antiferromagnets (this was explained in
Ref.~\onlinecite{4spin} merely by a footnote).
And we check its predictions on the eigenenergy clustering
patterns by carrying out
exact diagonalizations of the fcc Type I antiferromagnet. 
We will also correct some numerical mistakes
in a table in Ref.~\onlinecite{4spin}, resulting in
a much better agreement between diagonalization
and analytical results for the triangular lattice.

We start with a general Heisenberg Hamiltonian
  \begin{equation}
  \Hop(\{\ss_i\}) =  
  \frac{1}{2}\sum_{ij}J_{ij}\,\ss_i\cdot\ss_j,
  \label{eq-Ham}
  \end{equation}
where $i$ and $j$ run over $N$ lattice sites 
and each $\ss_i$ has quantum spin $s$.
$J_{ij}$ is the coupling constant between the sites $i$ 
and $j$ and is equal to $J_m$ when $i$ and $j$ are
$m$-th nearest neighbors ($m=1,2,...$). 

We assume that the ground state of $\Hop$ has long-range
antiferromagnetic order (of the spin directions).
In the simplest antiferromagnets (such as the 
square or triangular lattice with nearest-neighbor
interactions), the classical ground state is unique (modulo
rotations) and the quantum ground state has the same 
ordering pattern.
The antiferromagnets of interest here have 
continuous degeneracies -- not due to symmetry --
which are typically broken by quantum fluctuations.~\cite{hen89,FN-notkag}

The present work 
applies to a subset of all classically degenerate 
Heisenberg antiferromagnets, 
which we call ``$p$-sublattice.''
This means the lattice
divides into $p$ equivalent sublattices,  and the
classical ground states are precisely those in which 
(i) the spins are parallel within each sublattice;
(ii) the vector sum of all the sublattice magnetizations
is zero. The two main examples that belong
to this class and have motivated this study
are the triangular and fcc Type-I systems with
appropriate $J_1$ and $J_2$ interactions.
The $p$-sublattice antiferromagnets are
discussed in more detail below (Sec.~\ref{sec-psublattice}).

The effective Hamiltonian $\Heff$ is a
function of the $p$ sublattice spins.
The simplest terms of the required symmetry 
are of biquadratic form, and any more complicated
form would be difficult to use in an analytic calculation.
This effective Hamiltonian is derived by expanding around mean-field theory. 
That is, one invents a ``mean-field Hamiltonian''
$\HMF$ which has the same classical ground state as the
true one, and which has an exactly known quantum ground state. 
The full Hamiltonian is then written as 
   \begin{equation}
        \Hop \equiv \HMF + \delta \Hop
   \label{eq-Hsplit}
   \end{equation}
and $\Heff$ is obtained by second-order
perturbation theory in $\delta \Hop$. 

Mean-field theory may be set up in two different ways
and these produce two different recipes for $\Heff$. 
The first approach was followed by Larson and Henley~\cite{larson},
generalizing a formula of Long~\cite{long}. There, 
$\HMF$ consists of local fields that fix each spin
in its classical direction;
the result is a derivation of $\Heff$, usable for 
any antiferromagnet with a nontrivial degeneracies, but only
for an infinite system.
In this paper, we follow the second approach
to mean-field theory: that is, the zeroth-order
Hamiltonian is an infinite-range model
in which the individual spins and their products
are replaced by the sublattice averages.
This has the advantage that the finite-$N$ effects
are accounted for (this is important for 
comparing with exact-diagonalization results).

The calculation proceeds in several stages.
The first stage is to map 
each sublattice in the $N$-spin system approximately
to just one spin of maximum length.~\cite{lech95}
The net energy from the neglected spin fluctuations 
of the spins in each sublattice (away from the perfectly
aligned state), 
is approximated by an effective biquadratic interaction 
favoring collinear states.~\cite{hen89,she82}
Such interactions appeared first in the theory of selection among 
degenerate classical ground states 
(the so-called ``order-by-disorder'' effect).~\cite{hen89,larson,jacobs}

Mathematically, our calculation involves two steps. 
First, operator equivalents that operate
on sublattice spin states are obtained for
one spin and the product of two spins. This step is 
purely mathematical and follows from 
the Wigner-Eckart theorem concerning
the addition of angular momenta. The second step
is summing the operator equivalents over the lattice,
and this step depends on the geometric relationship 
of the antiferromagnet's sublattices. 

In this paper, we derive the
biquadratic effective Hamiltonian 
and apply it to the fcc Type I antiferromagnet (with nearest-
and next-nearest-neighbor interactions). 
We generalize this derivation 
to treat arbitrary (translation-invariant)
exchange couplings, in particular to properly handle
couplings which connect two spins of the same sublattice
(such as $J_2$ in the fcc Type I antiferromagnet).
Combining this with the remaining results of Ref.~\onlinecite{4spin}
on Bohr-Sommerfeld semiclassical quantization and
geometrical phase,
we find that the expected splitting pattern for 
the fcc Type I indeed agrees with numerical diagonalization
results for $N=32$ sites. (See Ref.~\onlinecite{LefRis00}
for a more detailed description of numerical diagonalization of
fcc systems and Ref.~\onlinecite{houle99} for an earlier 
study of another spin system, a three-spin cluster, using
many of the key semiclassical ideas here.)

This paper is organized as follows: we first introduce
the class of $p$-sublattice
antiferromagnets and give some examples (Sec.~\ref{sec-psublattice});
we then motivate and give the mean-field Hamitonian
(Sec.~\ref{sec-1storder});
next, the biquadratic effective Hamiltonian is derived
from second-order perturbation theory (this is the core
of the paper) (Sec.~\ref{sec-2ndorder}). We then apply
semiclassical considerations
to the resulting effective Hamiltonian (Sec.~\ref{sec-semicl})
analytically calculating
the energy-level pattern. The analytic result for the
type-I fcc antiferromagnet, as well
as a convenient four-spin exact diagonalization based
on the same effective Hamiltonian, is compared with
exact diagonalization of a 32-site system (Sec.~\ref{sec-num});
finally, we make some closing observations (Sec.~\ref{sec-discussion}).

\section {The $p$-sublattice antiferromagnets}
\label{sec-psublattice}

We will set up a formalism that includes all realizations
of $p$-sublattice antiferromagnets. 
It is convenient to define a geometric factor
$\zab$, as the number
of $m$-th neighbors on sublattice $\alpha$ of
a site on sublattice $\beta$.
(We will consistently use Greek indices $\alpha, \beta$
to label sublattices and $\mu, \nu, \rho$ to label
spin components.)
Notice that $\zab$ is not an attribute of the
Hamiltonian (\ref{eq-Ham}), but of the 
geometrical arrangement of sublattices
within the system.

Unless specifically noted, we will assume (and need)
{\it sublattice-pair permutation symmetry}.
This means that if we perform {\it any} of the
$p!$ permutations of the sublattice labels, the
geometric relationship between  any {\it pair}
$(\alpha, \beta)$ will be unchanged (modulo
lattice symmetries such are translations, rotations, 
and reflections).

The classical ground state manifold is
labeled by the unit vectors $\{ \mm_\alpha \}, \alpha =1, \ldots, p$
for the directions of the sublattices. 
Since we seek a {\it quantum-mechanical} Hamiltonian, 
however, it should be expressed in terms of the
corresponding operators, which (apart from length normalization)
are the sublattice spins, 
   \begin{equation}
   \SS_{\alpha}=\sum_{i:\alpha(i)=\alpha}\ss_i,
   \end{equation}
where $\alpha(i)=1,2,...,p$ gives the sublattice index of 
the spin $\ss_i$. Here the sum is over $\Nsub \equiv N/p$ sites
within the sublattice $\alpha$. 
Each $\SS_{\alpha}$ can be in states with total
spin ranging from zero to $\Ssub \equiv \Nsub s$. 
The classical state corresponds to a coherent state in
which every $\ss_i$ in sublattice $\alpha$ is aligned with $\mm_\alpha$, 
hence each sublattice spin has its maximum length $\Ssub$. 

Our final goal will be an effective Hamiltonian $\Heff$ defined on the
sub-Hilbert-space of ``$p$-sublattice states'', being all
states in which every sublattice spin has maximum length, 
$\SS_\alpha^2 = \Ssub (\Ssub +1)$;
thus $\Heff$ will have the form of a spin Hamiltonian. 
(Note that these ``$p$-sublattice states'' have maximum
{\em length}. The sublattice magnetization is
not determined and will be decided by diagonalizating
the secular Hamiltonian matrix as developed in the 
degenerate perturbation below.)
Along the way, we will often deal 
with the Hilbert space of one particular sublattice -- 
e.g. the operator equivalents are computed in this space. 
In particular, we let $|\pp_{\alpha}\rangle$
mean any eigenstate with $\SS_{\alpha}^2 = \Ssub(\Ssub+1)$
(maximum length) and simultaneously 
a definite $S_{\alpha}^z$.
The direct products of such states make up the
basis for $p$-sublattice states.  
On the other hand, $|\psii_{\alpha}\rangle$ 
will be used for a simultaneous eigenstate of 
$\SS_{\alpha}^2$ and $S_{\alpha}^z$
with {\it any} total spin,  $0\le S_\alpha\le \Ssub$.


The chief concrete examples have $p=4$, being:
(i) the $J_1$-$J_2$ triangular antiferromagnet,~\cite{lech95,chu92}  with
$J_1>0$ and second-neighbor coupling $J_2\in (J_1/8,J_1)$ --
the effective
Hamiltonian for this case was announced in Ref.~\onlinecite{4spin};
(ii) the Type I fcc antiferromagnet~\cite{og85,hen87} with 
$J_1>0, J_2 <0$. 
In either example, there are  only nearest and next-nearest
neighbor interactions ($J_m=0$ for $m>2$);
the Type I fcc differs from the triangular case in that
the next-nearest-neighbor couplings ($z^2$) 
connects sites of the {\it same} sublattice. 
(Table~\ref{t-Heffparams}, below, 
includes the $\zab$ values for these lattices.) 
The honeycomb and pyrochlore lattices 
also support four-sublattice states
for appropriately chosen exchange couplings ($J_1$ and $J_2$).
After allowing for rotations, 
a 4-sublattice antiferromagnet
still has a two-parameter family 
of classical ground states.~\cite{chu92,og85,hen87}

\NOTE {There is an interesting exotic triangular model
with 4-spin interactions, 
intended to represent the magnetic
behavior of a solid $^3$He monolayer on graphite, and 
which has a classical ground state of the 
tetrahedral type 
(T. Momoi, K.~Kubo, and K.~Niki,
Phys. Rev. Lett. 79, 2081 (1997))
However, this Hamiltonian would place
the collinear states quite high up, so one wouldn't see the near
degeneracy found here.  If one took the {\it degenerate}
4-sublattice state favored by $J_1$ and $J_2$, and added to 
the Hamiltonian a bit
of the 4-spin Hamiltonian, then one might well find a low-energy
subspace with a level sequence in the reverse of the order described
in this paper (since -- if the extra term is strong enough to beat the
collinearity term -- we favor the T rather than the C state.)}

The case $p=3$ is less interesting: 
there is just one way to add up three equal-length spins to make
total spin zero, so the classical or ``mean-field''
ground state is already unique.
\NOTE {Our effective Hamiltonian, which is only, 
defined within the ground state manifold, would be equivalent to a constant.
As for excited states, there is in this case no reason to expect 
each sublattice to behave approximately as if the spins were rigidly
coupled to be parallel.}
But our approach would be applicable to
the three-sublattice antiferromagnet in
an external field, in which case the ground state
{\it does} have nontrivial degeneracy.

There exist $p$-sublattice antiferromagnets 
with $p>4$, and possessing  
``sublattice-pair permutation symmetry'', 
although their Hamiltonians are physically improbable. 
For example, the $p=5$ case is realized 
by the $\sqrt 5 \times \sqrt 5$ sublattice;
in a square lattice without reflection symmetry.
The ground state has five sublattices, {\it e.g.,}
when the couplings to the $\pm [2,1]$ and $\pm [1, -2]$
fourth neighbors are strongly ferromagnetic, and the
nearest neighbor couplings are antiferromagnetic.
\NOTE{a site has one nearest neighbors from each of the
other four sublattices, and this symmetry 
holds for every class of neighbor vectors.}
A similar construction with $p=7$
may be made using the $\sqrt 7 \times \sqrt 7$ sublattice
of a triangular lattice.

\NOTE {We have not yet succeeeded in
finding a set of interactions
with the full symmetry of the lattice, such that
the only classical ground states are
these 5-sublattice states.  
For a generic set of $\{ J_m \}$, one can always
find ground states by the ``Luttinger-Tisza'' method.
Namely, on a Bravais lattice, 
the ordering wavevector $\QQ$ is one of those
which minimize 
     $J(\QQ) = \sum _{\rr} J(\rr) e^{i\QQ\cdot \rr},$
where $J(\rr)$ is the coupling of sites that differ by 
a displacement $\rr$.
Having found $\QQ$ by minimizing $J(\QQ)$, a valid
set of classical spin directions is the spiral state
${\bf s}(\rr) = (\cos \QQ\cdot \rr, \sin \QQ\cdot \rr, 0)$.
If $\QQ$ is a reciprocal lattice vector of the 
$\sqrt 5 \times \sqrt 5$ superlattice, then 
this is one of the 5-sublattice states. (Obviously
the five spin directions must sum to zero, since the
spiral state has zero total magnetization.)
In that case, in view of the sublattice-pair
permutation symmetry, all other 5-sublattice states
must have the same energy and hence are also 
classical ground states.  But there may be a much larger 
class of states which is degenerate, too.}

\section{Mean-field Hamiltonian}
\label{sec-1storder}

The objective of this section is to construct from $\Hop$ 
of Eq.~(\ref{eq-Ham}) a ``mean-field''
Hamiltonian $\HMF$ that is defined on the entire Hilbert space but
expressible in terms of the sublattice spins $\{ \SS_\alpha \}$.
In other words, it is a projection of the exact Hamiltonian 
onto a kind of infinite-range Hamiltonian. 
We will show that the ground states of $\HMF$ are all the
$p$-sublattice states; this high degeneracy corresponds to
the classical degeneracy. 
The real value of $\HMF$ is that we can split the
Hamiltonian in form (\ref{eq-Hsplit}) 
and obtain $\Heff$
via second-order perturbation in $\delta \Hop$ 
(see Sec.~\ref{sec-2ndorder}). 

The eigenstates of the lattice system that correspond one-to-one
to those of our $p$-spin effective Hamiltonian are not literally 
superpositions of the ``$p$-sublattice states.'' 
The true eigenstates include quantum fluctuations that
admix other states (as is implicit in our second-order perturbation), 
so the overlap with the $p$-sublattice subspace must decrease
exponentially with increasing system size $N$; indeed
the authors of Ref.~\onlinecite{lech95} computed the overlap
and found it small.~\cite{lhu}
The low-energy states of the lattice antiferromagnet are
related to the $p$-spin system's eigenstates much as 
the excited states of a Fermi liquid are related to 
those of noninteracting electrons.~\cite{FN-fermiliquid}
In this paper, our main interest is in analyzing the
clustering patterns of the low-lying energy spectrum.
We will focus on calculating eigenenergies, 
not magnetization expectations.

The mean-field Hamiltonian $\HMF$, to be a function only of
the sublattice totals $\{ \SS_\alpha \}$, 
must be symmetric under any
way of permuting the spins within a sublattice. 
Furthermore, the {\it states} that are symmetric under
such permutations are just the $p$-sublattice states, 
and $\HMF$ ought to have the same expectation as 
$\Hop$ in such a state. (This will be demonstrated in
the next section.)

The unique way to project $\Hop$ onto $\HMF$ and
satisfy these conditions 
is to average each term $\ss_i\cdot\ss_j$
over permutations of 
the sublattice(s) to which the sites $i$ and $j$ belong.
There are two cases. First, the two spins are on
different sublattices ($\alpha(i)\ne\alpha(j)$);
then we simply replace $\ss_i\cdot\ss_j \to
(\SS_{\alpha(i)}/\Nsub)\cdot(\SS_{\alpha(j)}/\Nsub)$.
Second, the spins are on the same sublattice 
($\alpha(i)=\alpha(j)=\alpha$ with $i\neq j$); in this case
there are $\Nsub(\Nsub-1)$ pairs of such scalar products
within one sublattice. (No permutation can ever make $i=j$.)
We use the identity
$\SS_\alpha^2=\Nsub s(s+1)+\sum_{i\ne j} \ss_i\cdot\ss_j$ to obtain
that $\ss_i\cdot\ss_j \to 
(\SS_\alpha^2-\Nsub s(s+1))/\Nsub(\Nsub-1)$
 in this case. 
(When $S_\alpha=\Ssub$, i.e., maximum length, this average
is simply $s^2$; restricted to the maximum-spin subspace, 
the projection we are doing here agrees with the operator equivalent
of $\ss_i\cdot \ss_j$, defined in (\ref{eq-scalarop}), below.)
Finally, putting these two kinds of terms
together and performing the sum, we get
   \begin{eqnarray}
   \nonumber
   \HMF&=&\frac{1}{2}\sum_{\alpha\ne\beta}\tJ_{\alpha\beta}
   \SS_\alpha\cdot\SS_\beta\\
   &+&\frac{1}{2}\sum_{\alpha}\tilde{J}_{\alpha\alpha}
   \frac{\Nsub}{\Nsub-1}\left(\SS_\alpha^2-\Nsub s(s+1)\right),
   \label{eq-H1}
   \end{eqnarray}
where 
   \begin{equation}
   \tJ_{\alpha\beta} \equiv \sum_m(J_m\zab)/\Nsub .
   \end{equation}

We restrict our attention to the class of
$p$-sublattice antiferromagnets with 
sublattice-pair permutation symmetry, 
so $z^m_{\alpha\beta}=z^m_{12}$ for each $m$ and $\alpha\ne\beta$ and
$z^m_{\alpha\alpha}=z^m_{11}$.
Thus $\tJ_{\alpha\beta}=\tJ_{12}$ 
for all $\alpha\ne\beta$ and $\tJ_{\alpha\alpha}=\tJ_{11}$
for all $\alpha$'s. The mean-field Hamiltonian becomes,
   \begin{eqnarray}
   \nonumber
   \HMF&=&\frac{1}{2}\tilde{J}_{12}\left(
   \SS_{\rm tot}^2-\sum_\alpha \SS_\alpha^2\right)\\
   &&+\frac{1}{2}\tJ_{11} \frac{\Nsub}{\Nsub-1}
   \left(\sum_\alpha\SS_\alpha^2-p\Nsub s(s+1)\right)\\
   &=&\frac{1}{2}\tilde{J}_{12}\SS_{\rm tot}^2
   -\frac{1}{2}\tJ_S
   \sum_\alpha\SS_\alpha^2
   -\frac{1}{2}p\tJ_{11}\frac{\Nsub^2}{\Nsub-1}s(s+1),
   \label{eq-H1simplified}
   \end{eqnarray}
where $\SS_{\rm tot}=\sum_\alpha\SS_\alpha$
and we have also defined
   \begin{equation}
   \tJ_S=\tJ_{12}-\frac{\Nsub}{\Nsub-1}\tJ_{11}.
   \label{eq-JS}
   \end{equation}

Next we consider the ground states for $\HMF$, 
which we denote $|\PP\rangle$. 
For this, we further
restrict to the situation where $\tJ_S>0$,
i.e., the coefficient in front of $\sum\SS_\alpha^2$ 
in (\ref{eq-H1simplified}) is positive.
(This condition is satisfied by both the triangular antiferromagnet
($\tJ_{11}=0$) and the fcc type-I case ($\tJ_{12}>0$ and $\tJ_{11}<0$).
See Table~\ref{t-Heffparams} for a collection of relevant
parameters.)

Then $|\PP\rangle$ should obviously satisfy $\SS_{\rm tot}^2 |\PP\rangle=0$
and $\SS_\alpha^2 |\PP\rangle = \Ssub(\Ssub+1)|\PP\rangle$ for 
every $\alpha$. 
In other words, the ground states of $\HMF$ are
just the $p$-sublattice states which are also total singlets,
so we call them ``$p$-sublattice singlet states.''
(A similar result is obtained for the classical ground state:
that each of the $p$ sublattice spin vectors 
is at maximum length and their vector sum is zero.) 
For $p=4$, there are $2\tilde{S}+1$ such 
singlets;~\cite{lech95} for $p=5$ and
$\Ssub$ is an integer, there are
$5\tilde{S}(\tilde{S}+1)/2+1$
(see Appendix~\ref{app-singlet}).
Lecheminant {\it et al}.~\cite{lech95}
already recognized that, in the
$J_1$-$J_2$ triangular antiferromagnet, 
the lowest energy singlets correspond to 
the four-sublattice singlet states.

We emphasize here that we have a degenerate family of singlets.
These states have the rotational symmetry as the true
finite-N ground state is known to have (i.e., they are
singlets). The sublattice spins have 
maximum {\em length} but the z components (magnetization expectations) 
are not determined.
Which linear combination of these states best represents
the ground state is only settled by the degenerate perturbation
calculation that follows, which diagonalizes an effective Hamiltonian
within this degenerate family of singlet states.

The ground state energy of (\ref{eq-H1simplified}) is
   \begin{equation}
      \EMF = \frac{1}{2}p\left[\tJ_{11}\Ssub^2-\tJ_{12} \Ssub(\Ssub+1)\right].
   \label{eq-E1}
   \end{equation}
Table~\ref{t-Heffparams} gives $\EMF$ for the
triangular and fcc lattices.

\NOTE{Parts of our derivation would probably apply in cases
where $\zab$ lacked the sublattice-pair permution symmetry, 
yet $\tJ_{\alpha\beta}$ accidentally possessed it.}


It is necessary to correct this mean-field Hamiltonian by $\Heff$, 
the effective biquadratic Hamiltonian,~\cite{4spin}
to predict the sequence of quantum numbers of the
low-lying states, and the clustering pattern of their eigenenergies.


\section{Biquadratic effective Hamiltonian}
\label{sec-2ndorder}

The calculation of the
second-order effective Hamiltonian $\Heff(\{\SS_\alpha\})$ will be presented 
now, and it is the central analytic result of this paper.
We will use the second-order degenerate perturbation theory \cite{LL}
for $\delta \Hop \equiv \Hop-\HMF$,
   \begin{equation}
   \langle\PP'|\Heff|\PP\rangle=-\sum_{\Psi}
   \frac{\langle\PP'|\delta \Hop|{\Psi}\rangle
   \langle{\Psi}|\delta \Hop|\PP\rangle}
   {E_{\Psi}-\EMF},
   \label{eq-deg}
   \end{equation}
where $|\PP\rangle$ and $|\PP'\rangle$ are both $p$-sublattice singlet
states (i.e.,  in the ground state manifold of $\HMF$) 
and $|\Psi\rangle$ is an excited state of $\HMF$.
In the process, we will also show that this choice of mean field
Hamiltonian enables the first-order correction to vanish, i.e.,
\begin{equation}
\label{eq-firstorder}
\langle\PP'|\dH|\PP\rangle=0.
\end{equation}

It is easy to show that
   \begin{equation}
   \label{eq-delH}
   \delta \Hop=\dHdiff+\dHsame,
   \end{equation}
where,
   \begin{eqnarray}
   \label{eq-dHdiff}
   && \dHdiff=\frac{1}{2}\sum_m J_m
   \sum_{{\tiny \begin{array}{c}\mth(ij)\\ \alpha(i)\ne\alpha(j)\end{array}}}
   \delta\ss_i\cdot\delta\ss_j  \nonumber\\
   && =\frac{1}{2}\sum_mJ_m\left(
   \sum_{{\tiny \begin{array}{c}\mth(ij)\\ \alpha(i)\ne\alpha(j)\end{array}}}
   \ss_i\cdot\ss_j
   -\frac{1}{\Nsub}\sum_{\alpha\ne\beta}\zab\SS_\alpha\cdot\SS_\beta\right),
   \end{eqnarray}
and,
   \begin{equation}
   \dHsame=\frac{1}{2}\sum_m J_m
   \sum_{{\tiny \begin{array}{c}\mth(ij)\\ \alpha(i)=\alpha(j)\\=\alpha
   \end{array}}}
   \left(\ss_i\cdot\ss_j-\frac{\SS_\alpha^2-\Nsub s(s+1)}
   {\Nsub(\Nsub-1)}\right),
   \label{eq-dHsame}
   \end{equation}
and $\delta\ss_i=\ss_i-\SS_{\alpha(i)}/\Nsub$.
Here the subscript ``diff'' means that the sum is over $(ij)$ pairs
on different sublattices and ``same'' means that they are on
the same sublattices; also, 
``$\mth (ij)$'' means a sum is restricted to pairs 
$(i,j)$ which are $m$-th neighbors. 
We should consider what kind of excited states
$|\Psi\rangle$ are present in the sum of the second-order 
perturbation theory (\ref{eq-deg}). Because this consideration
in part relies on the Wigner-Eckart theorem
and the concept operator equivalent, we will turn
to these mathematical concepts now.

\subsection {Operator equivalents}
\label{sec-opequiv}

The general problem of this sort of calculation is
to take an operator defined in a large Hilbert space, restrict 
this Hilbert space to a smaller one, and re-express its 
action by an operator acting on the restricted space. 
The latter operator will be called the ``operator equivalent''
of the first one.
In the present paper, we start from an operator 
$\hat{O}(\{\ss_i\})$ that acts on all states 
of the spins $\{\ss_i\}$ 
belonging to sublattice $\alpha$ and convert it 
into an operator $\hat{o}(\SS_\alpha)$ acting on states of a
single net sublattice spin $\SS_\alpha$.~\cite{FN-opeq}

The obvious tool for this calculation is the Wigner-Eckart
theorem \cite{wigner} which deals with angular momenta addition.
It says that among states of a single spin multiplet
(such as the maximum-spin states $|\phi_\alpha\rangle$), 
any operator that transforms as a 
particular representation of the rotation group, 
has matrix elements given by Clebsch-Gordan coefficients, 
apart from a single constant factor setting the overall scale. 
Hence we can replace that operator by a simpler one belonging 
to the same representation: here,  
a polynomial $\hat{o}(\SS_\alpha)$.
There is a unique such polynomial 
for each of the rotational symmetries 
(scalar, vector, or traceless-symmetric tensor). 
Specifically, we will be given operators $\hat{O}(\{\ss_i\})$, 
with a given symmetry, and write their matrix elements 
$\langle\phi'_\alpha|\hat{O}(\{\ss_i\})|\phi_\alpha\rangle
=c\langle\phi'_\alpha|\hat{o} (\SS_\alpha)|\phi_\alpha\rangle$
where $c$ is a constant that depends on $S_\alpha$, and possibly
on $i$, but not on the $z$-component of the spin 
(see e.g., Ref.~\onlinecite{GroupTheory}). 
In our calculation, we restrict our attention
to the maximum spin manifold $S_\alpha=\Ssub$
because it turns out that we will need only matrix elements
between the quantum $p$-sublattice singlet states $|\PP\rangle$
and $|\PP'\rangle$, where $S_\alpha=\Ssub$ for all $\alpha$;
in any case, 
when $\SS_\alpha < \Ssub$ there is more than one multiplet
and the Wigner-Eckart theorem is not enough to specify the
matrix elements. 

The results for scalar and vector operators are,
   \begin{eqnarray}
   \label{eq-scalarop}
   \langle\pp_\alpha'|\ss_i\cdot\ss_j|\pp_\alpha\rangle
     &=& s^2\langle\pp_\alpha'|\pp_\alpha\rangle, \nonumber\\
     && \quad \alpha(i)=\alpha(j)=\alpha, \quad i\ne j, \quad S_\alpha=\Ssub\\
   \label{eq-vectop}
   \langle\pp_\alpha'|s_i^\mu|\pp_\alpha\rangle
       &=&{1 \over \Nsub}
       \langle\pp_\alpha'|S_\alpha^\mu|\pp_\alpha\rangle, \nonumber\\
       && \quad \alpha(i)=\alpha, \quad S_\alpha=\Ssub.
   \end{eqnarray}
The constant coefficients, $c=s^2$ for the scalar operator
equivalent and $c=1/\Nsub$ for the vector operator equivalent,
were found by explicitly calculating the matrix elements
on both sides for a particular 
choice of states:  $|\phi_\alpha\rangle$
and $|\phi'_\alpha\rangle$ are both the state
with all spins aligned along $+z$, i.e., $S_\alpha^z=\Ssub$.
This is the most convenient choice to simplify the algebra. 

As noted above, every maximum-spin state $|\phi_\alpha\rangle$ 
is symmetric under 
permutations of the sites of sublattice $\alpha$, hence
our operator equivalent expressions
(\ref{eq-scalarop}) and (\ref{eq-vectop}) cannot depend on $i$ and $j$.
This symmetry could be used in place of the Wigner-Eckart theorem
to derive eqs.~(\ref{eq-scalarop}) and (\ref{eq-vectop}), but
only the Wigner-Eckart approach works in the tensor operator case 
(Sec.~\ref{sec-tensorop}, below).

\subsection {Energy denominators}

Now we are ready to compute the matrix elements
 $\langle \Psi|\delta \Hop|\PP\rangle$ and 
   denominators ${E_{\Psi}-\EMF}$ for the
excited states in the perturbation sum (\ref{eq-deg}). 
First we address $\dHsame$ which can be considered
as acting on states of sublattice $\alpha$, 
$|\pp_\alpha\rangle$ or $|\psi_\alpha\rangle$. 
Notice that the operator $\ss_i\cdot\ss_j$ 
with $\alpha(i)=\alpha(j)=\alpha$ commutes with $\SS_\alpha^2$. From 
(\ref{eq-dHsame}), 
therefore, $[\dHsame,\SS_\alpha^2]=0$ for all $\alpha$.
This means that, if the states $|\pp_\alpha\rangle$
and $|\psi_\alpha\rangle$ have different spin, 
the matrix element
$\langle\psi_\alpha|\dHsame|\pp_\alpha\rangle$ is zero.  
Since in our perturbation sum $|\PP\rangle$ is a maximum spin state,
the state $|\pp_\alpha\rangle$ has spin $\Ssub$
and so $S_\alpha=\Ssub$ for $|\psi_\alpha\rangle$ also.
Finally we have, for $\alpha(i)=\alpha(j)=\alpha$,
	\begin{eqnarray}
	\nonumber
	&&\langle\psi_\alpha|
	\ss_i\cdot\ss_j-\frac{\SS_\alpha^2-\Nsub s(s+1)}
   	{\Nsub(\Nsub-1)}|\pp_\alpha\rangle\\
	&&=\langle\psi_\alpha|\ss_i\cdot\ss_j-s^2|\pp_\alpha\rangle=0,
	\end{eqnarray}
where the last step relies on the scalar operator equivalent result
(\ref{eq-scalarop}). This tells that $\dHsame$
gives no contribution to the second-order perturbation (\ref{eq-deg}),
i.e., $\langle \Psi|\dHsame|\PP\rangle=0$.
Of course, the argument here also shows that $\dHsame$ gives no 
contribution to the first-order perturbation (\ref{eq-firstorder}),
i.e., $\langle \PP'|\dHsame|\PP\rangle=0$.

In the same way, we turn to the contribution by $\dHdiff$.
It is easy to show that $[(\delta\ss_i\cdot\delta\ss_j),\SS_{\rm tot}^2]=0$
which gives $[\dHdiff,\SS_{\rm tot}^2]=0$.
Since $|\PP\rangle$ has $\SS_{\rm tot}^2=0$, it follows 
that $\SS_{\rm tot}^2=0$ for $|\Psi\rangle$ also, 
otherwise $\langle\Psi|\dHdiff|\PP\rangle$ would be zero. 

Furthermore, that matrix element 
can be broken into terms of form 
$\langle \psi_1\psi_2\psi_3\psi_4|\delta\ss_i\cdot
\delta\ss_j|\pp_1\pp_2\pp_3\pp_4\rangle$. 
Assume without loss of generality that $\alpha(i)=1, \alpha(j)=2$, 
so the matrix element factors as 
$\langle \psi_1 |\delta\ss_i| \phi_1 \rangle \cdot 
 \langle \psi_2 |\delta\ss_j| \phi_2 \rangle 
 \langle \psi_3 | \phi_3 \rangle 
 \langle \psi_4 | \phi_4 \rangle $. 
Consider the first factor:
as defined above, $|\pp_1\rangle$ has
$S_1=\Ssub$, and the raising and lowering
operators in $\delta\ss_i$ can change $S_\alpha$ by
$+1$, $0$, or $-1$ only.
An increase $\Ssub\to \Ssub+1$ is obviously impossible
since $\Ssub$ is the maximum spin. 
Also, 
when $|\psi_1\rangle$ has the same length spin, 
$\Ssub$, 
the matrix element is zero by the construction 
$\delta\ss_i=\ss_i-\SS_{\alpha(i)}/\Nsub$, as found
using the operator equivalent (\ref{eq-vectop}), 
Hence, by elimination, the only states $|\psi_1\rangle$ 
that can give any nonzero matrix element
in $\langle \psi_1| \delta \ss_i | \phi_1\rangle$
must have spin $\Ssub-1$; the same holds for 
$|\psi_2\rangle$, whereas $|\psi_3\rangle$
and $|\psi_4\rangle$ obviously have the
same spin $\Ssub$ as do $|\phi_3\rangle$ 
and $|\phi_3\rangle$. 

Finally, every term of $\dHdiff$
contains products of two such operators $\delta\ss_i\cdot\delta\ss_j$,
so $\langle \Psi |\dHdiff |\PP\rangle$ is nonzero
only if $|\Psi\rangle$ has exactly two of the $p$
spins reduced by one. This gives
  \begin{eqnarray}
   \nonumber
   E_{\Psi}&=&-\frac{1}{2}\tJ_S((p-2)\Ssub(\Ssub+1)+2(\Ssub-1)\Ssub)\\
   &&\quad-\frac{1}{2}p\tJ_{11}\frac{\Nsub^2}{\Nsub-1}s(s+1).
  \end{eqnarray}
Subtracting the ground state
energy $\EMF$ (\ref{eq-E1}), the energy denominator 
in the perturbation sum (\ref{eq-deg}) that can give
nonzero contribution is always
  \begin{equation}
     E_{\Psi}-\EMF=2\tJ_S\Ssub
  \end{equation}

This argument also gives that
$\langle \PP'|\dHdiff|\PP\rangle=0$ within the ground states 
of $\HMF$. Together with the result 
$\langle \PP'|\dHsame|\PP\rangle=0$ obtained earlier in
this section, we conclude that the first-order correction
to $\HMF$ by $\dH$ is zero, i.e., eq.~(\ref{eq-firstorder})
is established.

\subsection{Reduction of Perturbation Sum}
\label{subsec-collapse}

We have now concluded that all the states $|\Psi\rangle$
that contribute nonzero numerator to the perturbation sum (\ref{eq-deg})
have the same denominator. This key fact enables
us to factor the denominators
out of sum (\ref{eq-deg}). 
Then we can extend the sum to run over {\it all}
states $|\Psi\rangle$, not just the excited ones.
(Every added term is zero, 
since $|\Psi\rangle$ is a maximum-spin
state in the added terms, 
and we just showed that $\langle \Psi| \delta \Hop| \PP\rangle$
is always zero in that case.)
Equation (\ref{eq-deg}) then becomes
   \begin{equation}
   \langle\PP'|\Heff|\PP\rangle=-
   \frac{\langle\PP'|(\dHdiff)^2|\PP\rangle}
   {2\tJ_S\Ssub}.
   \label{eq-perturbH2}
   \end{equation}

So instead of trying to sum over excited states in
eq.~(\ref{eq-deg}),
our goal now will be to find the operator equivalent
of $(\dHdiff)^2$ as in eq.~(\ref{eq-perturbH2}). 
Squaring (\ref{eq-dHdiff}) gives
   \begin{equation}
   \label{eq-dHdiff2}
   \langle\PP'|(\dHdiff)^2|\PP\rangle
   =\langle\PP'|\frac{1}{4}\sum_{mn}J_mJ_n \Oop_{mn}|\PP\rangle,
   \end{equation}
where,
   \begin{eqnarray}
   \label{eq-Hmn}
   && \Oop_{mn}=
   \sum_{{\tiny \begin{array}{c}\mth(ij)\\ \alpha(i)\ne\alpha(j)\end{array}}}
   \sum_{{\tiny \begin{array}{c}\nth(kl)\\ \alpha(k)\ne\alpha(l)\end{array}}}
   \Biggl[ (\ss_i\cdot\ss_j)(\ss_k\cdot\ss_l) \nonumber\\
   && -{1\over \Nsub^2}
   \left(\sum_{\alpha\ne\beta}\SS_\alpha\cdot\SS_\beta\right)
   \left(\sum_{\gamma\ne\delta}\SS_\gamma\cdot\SS_\delta\right)\Biggr] .
   \end{eqnarray}
In deriving this, we have used another operator equivalent relation
   \begin{equation}
   \langle\pp_\alpha'|s_i^\mu S_\alpha^\nu|\pp_\alpha\rangle
   ={1\over \Nsub}
   \langle\pp_\alpha'|S_\alpha^\mu S_\alpha^\nu|\pp_\alpha\rangle,
   \quad\alpha(i)=\alpha,
   \label{eq-sSequiv}
   \end{equation}
which is true because $|\phi_\alpha\rangle$, 
being a state of maximum $S_\alpha$, 
has the symmetry of permutation among
all sites $i$ in sublattice $\alpha$.

\subsection{Operator equivalent of quadratic operator}
\label{sec-tensorop}

To obtain $\HMF$, we have considered, in section \ref{sec-1storder}, 
the operator equivalents of scalar and vector operators. Here for $\Heff$, we
need the operator equivalent of $(\ss_i\cdot\ss_j)(\ss_k\cdot\ss_l)$
which can have two of the spin operators (e.g. $\ss_i$ and $\ss_k$)
on the same sublattice.
For this case we will need the operator equivalents for tensor operators.

As before, to find the operator equivalent we consider a
state of just one sublattice
$|\pp_\alpha\rangle$ 
with the maximum spin $\Ssub$.  
Recall the decomposition of a direct product of
two vector operators:
   \begin{equation}
   s_i^\mu s_k^\nu={1\over 3}\Scalar\delta_{\mu\nu}+\epsilon_{\mu\nu\rho}
   \Vector_{\rho}+\Tensor_{\mu\nu},
   \label{eq-sstensors}
   \end{equation}
where
   \begin{eqnarray}
   \Scalar &=&\ss_i\cdot \ss_k,\quad
   \Vector_\rho=
       {1\over 2}\epsilon_{\mu\nu\rho}\,s_i^\mu s_k^\nu,\quad
       \nonumber\\
   \Tensor_{\mu\nu}&=&
       {1 \over 2}(s_i^\mu s_k^\nu+s_i^\nu s_k^\mu)
        -{1\over 3}\delta_{\mu\nu}\Scalar.
   \label{eq-3tensors}
   \end{eqnarray}
$\Scalar$ is a scalar operator, $\Vector_\rho$ a vector operator, 
and $\Tensor_{\mu\nu}$
a second-rank symmetric traceless tensor operator.

By the Wigner-Eckart theorem as stated in Sec.~\ref{sec-opequiv}, 
the operator equivalents of the three operators
in (\ref{eq-3tensors}) are known functions
of $\SS_\alpha$, apart from coefficients
depending only on $(i,k)$.
As before, we obtain the coefficients
using the most convenient choice of states 
$S_\alpha^z=\Ssub$ (all spins aligned up) to simplify the algebra.
With $\alpha(i)=\alpha(k)=\alpha$, for
the scalar operator $\Scalar$, we obtain
   \begin{equation}
   \langle\pp_\alpha'|\Scalar|\pp_\alpha\rangle
   =(s^2+\delta_{ik}s)\langle\pp_\alpha'|\pp_\alpha\rangle.
   \end{equation}
Note we can have $i=k$ here,  because $i$ and $k$ come from two separate
scalar products $(\ss_i\cdot\ss_j)(\ss_k\cdot\ss_l)$, 
whereas earlier in (\ref{eq-scalarop}) we necessarily had $i\ne j$.
Here $\delta_{ik}$ appears because when $i=k$ 
we have $\ss_i^2=s(s+1)$. Following the same procedure for
the vector and tensor operators (using all-spins-aligned states
to compute coefficients), we get,
\begin{eqnarray}
&&\langle\pp_\alpha'|\Vector_\rho|\pp_\alpha\rangle
=\delta_{ik}\frac{i}{2\Nsub}\langle\pp_\alpha'|S_\alpha^\rho
|\pp_\alpha\rangle,   \\
\nonumber
&&\langle\pp_\alpha'|\Tensor_{\mu\nu}|\pp_\alpha\rangle
=\frac{1}{\Nsub}\frac{2s-\delta_{ij}}{2\Ssub-1} \\
&&\quad\langle\pp_\alpha'|
\Bigl[{1 \over 2}(S_\alpha^\mu S_\alpha^\nu+S_\alpha^\nu S_\alpha^\mu)
   -{1 \over 3}\delta_{\mu\nu}\Ssub(\Ssub+1)\Bigr]
|\pp_\alpha\rangle.
\end{eqnarray}
And putting the three terms together as in (\ref{eq-sstensors}), we get,
   \begin{eqnarray}
   \nonumber
   &&\langle\pp_\alpha'|s_i^\mu s_k^\nu|\pp_\alpha\rangle=
   \langle\pp_\alpha'|\Bigl\{{1 \over 3}(s^2+\delta_{ik}s)\delta_{\mu\nu}
    + \delta_{ik}{i \over 2\Nsub}\epsilon_{\mu\nu\rho} S_\alpha^\rho\\
   &&\quad+{1 \over \Nsub}\frac{2s-\delta_{ik}}{2\Ssub-1}
   \Bigl[{1 \over 2}(S_\alpha^\mu S_\alpha^\nu+S_\alpha^\nu S_\alpha^\mu)
   -{1 \over 3}\delta_{\mu\nu}\Ssub(\Ssub+1)\Bigr]
   \Bigr\}|\pp_\alpha\rangle,
    \nonumber\\
   && \qquad\alpha(i)=\alpha(k)=\alpha.
   \label{eq-tensorop}
   \end{eqnarray}
It may be checked
that summing both sides of (\ref{eq-tensorop}) indeed 
gives (\ref{eq-sSequiv}).
As another check, notice that when $i$ and $k$
are distinct sites belonging to the same sublattice
$\alpha$, the spin operators $s_i^\mu$   and $s_k^\nu$ commute.
Hence the operator equivalent of $s_i^\mu s_k^\nu$
must be symmetric under exchange of $i\leftrightarrow k$, 
even though $[ S_{\alpha}^\mu, S_{\alpha}^\nu]\neq 0$;
indeed (\ref{eq-tensorop}) has this property.

We return to consider matrix elements between 
$p$-sublattice states such as $|\PP\rangle$
and $|\PP'\rangle$.
To compute the effective Hamiltonian due to $(\dHdiff)^2$, 
we will need the operator equivalent of $(\ss_i\cdot\ss_j)(\ss_k\cdot\ss_l)$
(see (\ref{eq-Hmn})). We give here the representatives of the three 
possible situations. (All others are equivalent, by
permutation of the sublattice indices.)

\subsubsection{Case $\alpha(i)=1,\alpha(j)=2,\alpha(k)=3,\alpha(l)=4$}
\label{case-1}
   \begin{equation}
   \label{eq-1234}
   \langle\pp'|
   (\ss_i\cdot\ss_j)(\ss_k\cdot\ss_l)
   |\pp\rangle
   =
   \langle\pp'|
   {1\over\Nsub^4}(\SS_1\cdot\SS_2)(\SS_3\cdot\SS_4)
   |\pp\rangle.
   \label{eq-nosame}
   \end{equation}

\subsubsection{ Case $\alpha(i)=\alpha(k)=1,\alpha(j)=2,\alpha(l)=3$}
\label{case-2}
   \begin{eqnarray}
   \label{eq-1213}
   &&\langle\pp'|(\ss_i\cdot\ss_j)(\ss_k\cdot\ss_l)|\pp\rangle
   =\langle\pp'|\Bigl\{
   \frac{s^2}{3\Nsub^2}(\SS_2\cdot\SS_3)
    \nonumber\\ 
   &&+2s\, G(\SS_1,\SS_2,\SS_3) 
   +\Bigl[\frac{s}{3\Nsub^2}(\SS_2\cdot\SS_3)-
   G(\SS_1,\SS_2,\SS_3) 
    \nonumber\\ 
   &&+\frac{i}{2\Nsub^3}\SS_1\cdot(\SS_2\times\SS_3)\Bigr]\delta_{ik}
   \Bigr\}|\pp\rangle,
   \end{eqnarray}
where
   \begin{eqnarray}
   &&G(\SS_1,\SS_2,\SS_3)=
   \frac{1}{\Nsub^3(2\Ssub-1)}
   \Bigl[\frac{1}{2}\bigl((\SS_1\cdot\SS_2)(\SS_1\cdot\SS_3)
    \nonumber\\ 
    &&+(\SS_1\cdot \SS_3)
    (\SS_1\cdot\SS_2)\bigr)-\frac{1}{3}\Ssub(\Ssub+1)\SS_2\cdot\SS_3
   \Bigr].
   \end{eqnarray}

\subsubsection{Case $\alpha(i)=\alpha(k)=1,\alpha(j)=\alpha(l)=2$}
\label{case-3}
   \begin{eqnarray}
   \nonumber
   &&\langle\pp'|(\ss_i\cdot\ss_j)(\ss_k\cdot\ss_l)|\pp\rangle
   =
   \langle\pp'|\Bigl\{
   {1\over 3}s^4+4s^2\,
   F(\SS_1,\SS_2)\\
   \nonumber
   &&+\Bigl[{1\over 3}s^3-2s\,
   F(\SS_1,\SS_2)
   \Bigr](\delta_{ik}+\delta_{jl})\\
   \label{eq-1212}
   &&+\Bigl[{1\over 3}s^2+
   F(\SS_1,\SS_2)
   -\frac{1}{2\Nsub^2}(\SS_1\cdot\SS_2)
   \Bigr](\delta_{ik}\delta_{jl})
   \Bigr\}|\pp\rangle,
   \end{eqnarray}
where,
   \begin{eqnarray}
   \label{eq-F}
   F(\SS_1,\SS_2)&=&\frac{1}{\Nsub^2(2\Ssub-1)^2}
   \Bigl[(\SS_1\cdot\SS_2)^2+
    \nonumber\\
   && {1\over 2}(\SS_1\cdot\SS_2)-
   {1\over 3}\Ssub^2(\Ssub+1)^2\Bigr].
   \end{eqnarray}

These three formulas are obtained from summing over $\mu,\nu$
in $\sum (\ss_i^\mu\ss_j^\mu)(\ss_k^\nu\ss_l^\nu)$ and using
(\ref{eq-vectop}) and (\ref{eq-tensorop}) for the operator equivalents 
of one single spin and the product of two spins on the same sublattice.

\subsection{Summing over the lattice}

With the operator equivalent of $(\ss_i\cdot\ss_j)(\ss_k\cdot\ss_l)$
we can sum over the lattice and obtain the operator equivalent of 
$\Oop_{mn}$
in (\ref{eq-Hmn}).
The results depend on 
how the sublattices are related geometrically to each other, 
via $\zab$ defined in Sec.~\ref{sec-psublattice}.
Again the representatives of the three possible
situations are the following.
\subsubsection {Case $\alpha(i)=1,\alpha(j)=2,\alpha(k)=3,\alpha(l)=4$}

Here we have $\Nsub z^m_{12}$ pair of $(\alpha(i)=1,\alpha(j)=2)$
and $\Nsub z^m_{34}$ pair of $(\alpha(k)=3,\alpha(l)=4)$ in the 
double sum $\sum_{\mth}\sum_{\nth}$. The contribution of (\ref{eq-1234})
to the total sum is
   \begin{equation}
   \frac{1}{\Nsub^2}z^m_{12}z^m_{34}(\SS_1\cdot\SS_2)(\SS_3\cdot\SS_4).
   \label{eq-1234total}
   \end{equation}

\subsubsection{Case $\alpha(i)=\alpha(k)=1,\alpha(j)=2,\alpha(l)=3$}

$\Nsub z^m_{12} \Nsub z^m_{13}$ terms 
and $\Nsub z^m_{12}  z^m_{13}$ of them have $i=k$ 
(for the $\delta_{ik}$ terms in (\ref{eq-1213})).
The contribution to the total sum turns out to be simple
   \begin{equation}
   \frac{1}{\Nsub^2}z^m_{12}z^m_{13}(\SS_1\cdot\SS_2)(\SS_1\cdot\SS_3).
   \label{eq-1213total}	
   \end{equation}
We can obtain this result either 
from the complicated operator
equivalent (\ref{eq-1213}) or from the fact that
here the double sums $\sum_{\mth}\sum_{\nth}$ can
be summed separately to give the sublattice spins.

\subsubsection{Case $\alpha(i)=\alpha(k)=1,\alpha(j)=\alpha(l)=2$}

This is the most complicated case. There are
$\Nsub z^m_{12} \Nsub z^m_{12}$ terms
and $\Nsub z^m_{12} z^m_{12}$ of them have $i=k$
(for the $\delta_{ik}$ term in (\ref{eq-1212})),
$\Nsub z^m_{12} z^m_{21}$ of them have $j=l$ (for $\delta_{ik}$ term),
and $\Nsub z^m_{12}$ of them have $i=k$ and $j=l$
(for $\delta_{ik}\delta_{jl}$ term).
Here the contribution to the total sum is
   \begin{eqnarray}
   &&\frac{1}{\Nsub^2}(z^m_{12})^2(\SS_1\cdot\SS_2)^2
     \nonumber \\
   &&-\frac{1}{\Nsub^2}(z^m_{12})^2
   \Biggl[\frac{1}{(2\Ssub-1)^2}(\SS_1\cdot\SS_2)^2
   -\frac{1}{2}\left(1-\frac{1}{(2\Ssub-1)^2}\right)
     \nonumber \\
    && (\SS_1\cdot\SS_2)
   +\frac{\Ssub^3(\Ssub-2)}{(2\Ssub-1)^2}\Biggr]
     \nonumber \\
   &&+\delta_{mn}\Nsub z^m_{12}
   \left(\frac{1}{3}s^2+\Nsub F(\SS_1,\SS_2)
   -\frac{1}{2\Nsub^2}(\SS_1\cdot\SS_2)\right),
   \label{eq-1212total}
   \end{eqnarray}
where $F$ is given above in (\ref{eq-F}).

\subsection{Final form of effective Hamiltonian}

Combining the three terms 
(\ref{eq-1234total},\ref{eq-1213total},\ref{eq-1212total}) 
and their cyclic permutations, we get
   \begin{eqnarray}
   \langle\pp'|\Oop_{mn}|\pp\rangle&=&
   \langle\pp'|\frac{2}{\Nsub^2(2\Ssub-1)^2}
   \sum_{\alpha\neq\beta}\Bigl[
   z^m_{\alpha\beta}(\Nsub\delta_{mn}-z^n_{\alpha\beta})
     \nonumber \\
   \bigl(({\SS_\alpha\cdot\SS_\beta})^2
   &-&2\Ssub(\Ssub-1)({\SS_\alpha\cdot\SS_\beta})+\Ssub^3(\Ssub-2)\bigr)
   \Bigr]|\pp\rangle
   \end{eqnarray}

Finally, we perform $\sum_{mn}$ in (\ref{eq-dHdiff2}) and 
use (\ref{eq-perturbH2}) 
and the fact that a singlet state $|\PP\rangle$ is simply a linear
combination of $p$-sublattice states $|\pp\rangle$
to obtain our result for the biquadratic effective Hamiltonian, 
   \begin{eqnarray}
   \label{eq-H2}
   \Heff&=&-\frac{K_{12}}{4\tJ_S\Nsub\Ssub(2\Ssub-1)^2}
   \sum_{\alpha\neq\beta}
   \Bigl[(\Sab)^2
     \nonumber \\
     &-& 2\Ssub(\Ssub-1)(\Sab)+\Ssub^3(\Ssub-2)\Bigr],
   \end{eqnarray}
where
   \begin{equation}
   K_{\alpha\beta}=\sum_m(J_m)^2 z^m_{\alpha\beta}
   -\frac{1}{\Nsub}\left(\sum_m J_m z^m_{\alpha\beta}\right)^2,
   \label{eq-K12}
   \end{equation}
and we have used permutation symmetry $K_{\alpha\beta}=K_{12}$.
($\Nsub$ and $\zab$ are defined in Sec.~\ref{sec-psublattice}.)
The middle term of (\ref{eq-H2}) can be simplified, since
it is linear in $\HMF$ which has the same value for any
of the $p$-sublattice singlets that we are considering.
The final result is 
   \begin{equation}
   \Heff(\{\SS_\alpha\})=-\tilde{K}\sum_{\alpha<\beta}(\Sab)^2+\tilde{C},
   \label{eq-H2simpler}
   \end{equation}
where 
	\begin{equation}
	\tilde{K}=\frac{K_{12}}{2\tJ_S\Nsub\Ssub(2\Ssub-1)^2},
	\end{equation}
and 
	\begin{equation}
	\tilde{C}=-\frac{1}{2}
	p\tilde{K}\Ssub^2((p+1)\Ssub^2-2(p-1)\Ssub-2).
	\end{equation}
By recalling that $\tJ_S \sim N^{-1}$ (see eq.~(\ref{eq-JS}))
and that $\SS_\alpha\sim\Ssub{}=Ns/p$, 
it may be checked that (\ref{eq-H2simpler}) scales with $N$. 


In the limit $N\to\infty$, the operator
$\SS_\alpha$ becomes perfectly defined and can 
be equated to a c-number, $\Ssub \mm_\alpha$, 
where $\mm_\alpha$ is a unit vector denoting the classical direction
of sublattice $\alpha$. 
Then our effective Hamiltonian (\ref{eq-H2simpler}) 
simplifies further to
	\begin{equation}
	-\frac{(\sum_m J_m^2 z_{12}^m)s}{8(\tJ_{12}-\tJ_{11})}
	\left(\sum_{\alpha<\beta}(\mm_\alpha\cdot\mm_\beta)^2+
	\frac{1}{2}p(p+1)\right).
	\label{eq-H2Ninf}
	\end{equation}

Earlier, Larson and Henley\cite{larson}
also obtained a second-order effective Hamiltonian 
starting from a different sort of mean-field theory:
	\begin{equation}
	\Delta \HH =-\frac{s^2}{8h_0}\sum_{\langle ij \rangle}J_{ij}^2
	(1-\mm_i\cdot\mm_j)^2.
	\label{eq-lh}
	\end{equation}
Here $\mm_i$ (called ``$\hat{z}_i$'' in Ref.~\onlinecite{larson})
is the classical spin unit vector at site $i$
and $h_0=|\sum_j J_{ij}s\mm_j|$ is the magnitude of the 
local field at $i$. 
Their result is valid for Hamiltonians more general than 
the ``$p$-sublattice'' category, on the other hand it is
valid only in infinite systems. 
In the special case
that sites are grouped into $p$ sublattices,
as assumed in this paper, it can be checked that (\ref{eq-lh})
does reduce to our effective Hamiltonian 
(\ref{eq-H2Ninf}) for the $N\to \infty$ limit,  
with $h_0=(\tJ_{12}-\tJ_{11})\Nsub s$.


\section{Semiclassical calculation of singlet energies}
\label{sec-semicl}

The effective biquadratic Hamiltonian (\ref{eq-H2simpler}) 
is the basis for an 
analytic approximation of the energy eigenvalues for a $p=4$ spin
antiferromagnet, as detailed in Ref.~\onlinecite{4spin}. 
A further series of mappings is applied to the effective 
Hamiltonian resulting in a semiclassical problem
of one spin with a quartic Hamiltonian.
Bohr-Sommerfeld quantization and tunneling considerations
are then applied to this one-spin problem to
obtain energies and level splittings.
The procedure here is identical to that in
Ref.~\onlinecite{4spin} which worked out the triangular lattice case, 
except that the coefficients are different, 
being functions of $J_1$ and $J_2$ as well
as the type and size of the lattice.

This section recapitulates and somewhat
extends the recent calculation by
two of us~\cite{4spin}
of the low-lying singlet energies
of the four-sublattice Heisenberg antiferromagnets, 
assuming long-range order in the $N\to \infty$ limit. 
That derivation started from a four-spin system with 
the biquadratic effective Hamiltonian (\ref{eq-H2simpler})
derived above. 
When the total spin is zero, as is the case here, 
the classical dynamics is separable
into a trivial and a nontrivial part 
(Sec.~\ref{subsec-singlespin}).
The nontrivial part is very similar to 
a single spin in an anisotropy field 
of cubic symmetry.~\cite{harter,robbins,pokr98}

Next, using Bohr-Sommerfeld quantization of the 
classical orbits of this effective one-spin Hamiltonian, 
the low-lying singlet energies were shown to form 
{\it clusters} of two or three degenerate levels and
the energies were approximately estimated (Sec.~\ref{subsec-BS}). 
It was found that the lower energy clusters are threefold
degenerate, corresponding to excitations near the ``collinear'' 
states of the classical ground state manifold (with
four spins forming two pairs that point
in opposite directions), and the
higher energy clusters are twofold degenerate, 
corresponding to noncollinear ``tetrahedral'' states
(with four spins pointing at the four vertices of a tetrahedron).

Finally, tunneling between Bohr-Sommerfeld orbits in 
different energy wells generates splittings
exponentially small (in the cluster size) within the level clusters, 
except when there is destructive interference of tunneling
amplitudes associated with different paths which leaves
the clusters unsplit.
From the geometrical phase of the tunneling paths,~\cite{geom}
one could predict the pattern of split and unsplit states in
the clusters (Sec.~\ref{subsec-cluster}).
The predicted pattern of level clustering 
was in accord with numerical
data\cite{lech95} for the $J_1$-$J_2$ triangular lattice with 
$N=16$ and $N=28$. 

We add a new discussion of the other quantum numbers, derived
from the spatial symmetries of the spin arrangements
(Sec.~\ref{subsec-symmetries}). Since it is advantageous
for exact diagonalizations to break the Hamiltonian 
into blocks with different symmetries, the symmetry
eigenvalues are often obtained as a byproduct, which
may now be checked against our predictions. 

The section concludes (Sec.~\ref{subsec-N})
with a discussion of the
system-size dependences of the various energy scales
characterizing low-lying eigenstates. 
This illustrates how brute-force diagonalization of
a (necessarily small) system can 
be misleading as to the energy scales of a large system, 
and how an analytic view can correct this.

\subsection{Coherent states and separation of rotational variables}
\label{subsec-singlespin}

We now carry out the semiclassical
calculation on our quantum $p$-spin effective Hamiltonian 
(\ref{eq-H2simpler}) as in Ref.~\onlinecite{4spin}. 
Given a quantum spin Hamiltonian, we can obtain 
upper and lower bounds of the quantum partition function
using the classical integral representation.~\cite{lieb}
A classical spin Hamiltonian can then be derived 
and is the starting point of semiclassical calculations.~\cite{shankar}
We note there are multiple ways to define this semiclassical 
spin Hamiltonian and two of which correspond to the upper 
and lower bounds of the partition function respectively
(they are called $\HH_Q$ and $\HH_P$ respectively 
in Ref.~\onlinecite{shankar}). Here we will use the $\HH_Q$ 
option which is obtained by taking expectations of the quantum
Hamiltonian within spin coherent states.

Using spin coherent states $|\{\mm_\alpha\}\rangle$ 
where $\mm_\alpha$ is a unit vector
(see, e.g., Ref.~\onlinecite{radcliffe}), with $1\le\alpha\le p$, we get,
\begin{eqnarray}
\nonumber
&&\Heffscalar(\{\mm_\alpha\})\equiv
\langle\{\mm_\alpha\}|\Heff|\{\mm_\alpha\}\rangle\\
&&=-K_U\sum_{\alpha<\beta}(\mm_\alpha\cdot\mm_\beta)^2+C_U,
\label{eq-H2semi}
\end{eqnarray}
where $K_U=\tilde{K}\Ssub^2(\Ssub-1/2)^2$
and $C_U=-p\tilde{K}\Ssub^2(4\Ssub(p-1)+p+1)/8+\tilde{C}$.
For a closed orbit in this space $\{\mm_\alpha(t)\}$ parametrized
by $t$, the geometrical phase (important for semiclassical calculation)
is,
   \begin{equation}
      \Phi  = \sum _{\alpha=1}^{4} \Ssub \Omega(\{\mm_\alpha(t)\}), 
   \label{eq-topo}
  \end{equation}
where $\Omega(\{\mm(t)\})$ denotes the spherical area the
trajectory $\mm(t)$ has swept out on the unit sphere
around the $z$ axis. Because each of the $p$ unit vectors
lives on a two-dimensional spherical surface, so
with the constraint $\sum\mm_\alpha=0$, we have
a $(2p-3)$-dimensional space specified by $\{\mm_\alpha\}$.
Furthermore, the global rotation separates from
the internal rotations (at least for $p=4$;~\cite{4spin}
this depends only on the permutation symmetry of
eq.~(\ref{eq-H2simpler}) ). 
We are left with a simpler problem on a two-dimensional spherical surface
specified by a new unit vector $\hat\nn$.

From now on, we will work on the special case $p=4$
which is the case for the fcc and triangular lattices.
First, let $\nn_\mu=(\mm_\mu+\mm_4)/2$, $\mu=1,2,3$
(not a unit vector). 
It is easy to check that these three vectors are
mutually orthogonal; therefore, given a coordinate system
$\{\ee_\mu\}$, we can rotate $\{\nn_\mu\}$, with
a proper rotation matrix $\RC$, so that
$\RC\nn_\mu=n_\mu\ee_\mu$, where $n_\mu$ are scalar
coefficients. We then form a new vector $\hat\nn=(n_1,n_2,n_3)$;
it is easy to check that this is a unit vector
($\nn_1^2+\nn_2^2+\nn_3^2=1$).
In terms of this $\hat\nn$, we obtain a one-spin semiclassical
Hamiltonian from (\ref{eq-H2semi}),
\begin{equation}
\Hone=-8K_U\sum_{\mu=1}^3 n_\mu^4+2K_U+C_U.
\label{eq-H2onespin}
\end{equation}
It is easy to show that $\RC\mm_1=(n_1,-n_2,-n_3)$,
$\RC\mm_2=(-n_1,n_2,-n_3)$, $\RC\mm_3=(-n_1,-n_2,n_3)$,
$\RC\mm_4=(n_1,n_2,n_3)$, which says that
each $\mm_\alpha(t)$ traces out the same shaped trajectory as
$\hat\nn(t)$ and makes the same contributation to 
the geometrical phase (\ref{eq-topo}). Therefore, we have
$\Phi=4\Ssub\Omega(\hat\nn(t))=S\Omega(\hat\nn(t))$, where
we have used $S=4\Ssub=N s$. In the sense of semiclassical
orbits and geometrical phases, we have mapped
the problem of Hamiltonian (\ref{eq-H2semi})
with four spins $\Ssub$ to the problem
of Hamiltonian (\ref{eq-H2onespin}) with one spin $S$.

It must be pointed out that the definition of our coherent
states $|\{\mm_\alpha\}\rangle$ included an averaging over
all rotations. It follows that the mapping
to $\hat\nn$ is not one-to-one. If $\RC_\pi$ is a $\pi$-rotation
about any coordinate axis $\ee_\alpha$, then 
$(\RC_\pi \RC) \nn_\mu ={\nn'}_\mu \ee_\mu$, and so
the rotation matrix $\RC_\pi\RC$ satisfies our condition
just as well as $\RC$. 
In other words, in the $\hat\nn$ representation, two vectors
related by a $\pi$ rotation about any coordinate axis represent
the same four-spin state $\{\mm_\alpha\}$.
We need to take this discrete redundancy into account
in the semiclassical calculation below.~\cite{FN-othersphere}

\subsection {EBK quantization}
\label{subsec-BS}

To obtain energies from the semiclassical spin Hamiltonian, we
use the EBK, or Bohr-Sommerfeld (BS), 
quantization condition which says that the geometrical phase
of the Bohr-Sommerfeld orbit is $2\pi l$ where $l=0,1,...,2S$.~\cite{FN-maslov}
(The BS semiclassical calculation described in this section
was carried out in Ref.~\onlinecite{4spin} for the
triangular system. The difference
here for the fcc system is the value of $S$.)
On the sphere, the spherical angle enclosed by 
two curves $\theta_1(\phi)$ and $\theta_2(\phi)$ 
is $\int(\cos\theta_1-\cos\theta_2)d\phi$,
where the integration limits are determined
by the geometrical locations of the two curves. 
For our quartic Hamiltonian, we first consider 
	\begin{equation}
	u=n_1^4+n_2^4+n_3^4=x^2+(1-x)^2 a(\phi),
	\end{equation}
where $x=\cos^2 \theta$ and $a(\phi)=\cos^4\phi+\sin^4\phi=(3+\cos 4\phi)/4$.
Solving this quadratic equation, we get,
	\begin{equation}
	\cos\theta=\pm\sqrt{\frac{a\pm\sqrt{u-(1-u)a}}{a+1}},
	\end{equation}
where the $\pm$ signs need to be chosen according 
to the location of the orbit.

As mentioned in Ref.~\onlinecite{4spin}, two types of orbits on the
$\hat\nn$ unit sphere are relevant. (1) ``Collinear'' (C) states
that orbit around the three coordinate axes (they are 
therefore three-fold degenerate). The quantization condition 
for these states is $\Phi=4\pi l$ (a factor of two
is present because of the rotation by $\pi$ redundancy mentioned above).
(2) ``Tetrahedral'' (T) states that orbit around two axes
$(\pm 1,1,1)/\sqrt{3}$ (and equivalents) 
(they are then two-fold degenerate).
Here the quantization condition is $\Phi=2\pi l$.

For the $C$-type orbits, the BS quantization condition
leads to the equation,
	\begin{equation}
	\Omega_{C_l}=
	\int_0^{2\pi}\left(1-\sqrt{\frac{a(\phi)+\sqrt{u-(1-u)a(\phi)}}
	{a(\phi)+1}}\right)
	d\phi=\frac{4\pi l}{S}.
        \label{eq-BSquantC}
	\end{equation}
For $T$-type orbits, we get,
	\begin{eqnarray}
	\nonumber
	\Omega_{T_l}=&&2\int_{1/2}^{u/(1-u)}\Bigl[
	\sqrt{\frac{a+\sqrt{u-(1-u)a}}{a+1}}\\
	&&-\sqrt{\frac{a-\sqrt{u-(1-u)a}}{a+1}}
	\Bigr]
	\frac{da}{\sqrt{1-(4a-3)^2}}=\frac{2\pi l}{S},
	\end{eqnarray}
where in the last integral we have changed integration variable to $a$,
using $d\phi=\pm da/\sqrt{1-(4a-3)^2}$ ($\pm$ signs determined by location).
Without loss of generality, the $C$-orbit equation is obtained from 
orbits around $(0,0,1)$ and the $T$-orbit equation from
those around $(1,1,1)/\sqrt{3}$.
For given $S$ and $l$,
these two equations are solved numerically for $u$ using Mathematica,
and the semiclassical energy is then $K_U(2-8u)+C_U$.
Semiclassical energies are presented in Tables~\ref{t-levels} 
and \ref{t-triangular} for fcc and triangular lattices respectively.
The result for the triangular case corrects
a mistake in the last column of Table~I in Ref.~\onlinecite{4spin}
which came from a wrong sign in the expression for $C_U$
just under eq.~(6) in Ref.~\onlinecite{4spin}. The correct expression
should be $C_U=-\Ssub^2(5\Ssub+5/2)\tilde{K}+\tilde{C}_{\rm biq}$.

\subsection{Tunnel splittings and eigenvalue fine structure}
\label{subsec-cluster}

The preceding section establishes that the low-lying levels
are grouped into clusters of three in the lower part 
(corresponding to the three-fold degenerate $C$-type orbits)
and clusters of two in the higher part 
(corresponding to the two-fold degenerate $T$-type orbits).
Average energies for the clusters have been calculated
from the \BS quantization condition.
Finer level splittings within each cluster are results of
tunneling between \BS orbits, and degeneracies
can remain due to destructive interference of 
the geometrical phases of the tunneling paths.~\cite{geom}
Using phase considerations, we can therefore obtain the 
split/unsplit pattern for energy clusters:
it depends on $N, s, l$.

We here simply quote the result from Ref.\onlinecite{4spin}. If we write
``$(\nu_1,\nu_2)$'' for the pattern of eigenvalues in a cluster, 
meaning that the (lower,higher) eigenvalues have degeneracies
$(\nu_1, \nu_2)$, respectively, then 
for the three-fold degenerate $C$-type orbits,
we obtain a cluster pattern $(2,1)$ if the expression $Ns/2-3l$ is odd 
or $(1,2)$ if it is even; and for the two-fold degenerate
$T$-type orbits, we get $(1,1)$ when $Ns-4l$ is divisible by three,
or $(2)$, i.e., unsplit, otherwise.~\cite {FN-patternerror}
Notice that, ignoring whether the splittings are large or small, 
the pattern is $11221122...$ at either end. Indeed this
sequence continues unbroken across the energy of the classical
separatrix of the orbits, which is the boundary between the two
clustering behaviors (i.e., where the $C$-type orbits meet the $T$-type).
As energy increases, the splitting within each cluster grows, 
until at the separatrix energy ($u=1/2$) 
it is typically
comparable to the separation between successive
clusters, causing an ambiguity in labeling the levels.

The case of fcc type-I with $N=16$ 
is trivial since $\delta \Hop \equiv 0$ which means
that the four-sublattice singlets are the {\it exact} ground states, 
and are exactly degenerate.
Our calculation, of course, gets this right since it gives
a zero coefficient in the effective Hamiltonian
($K_{12}=4(\Nsub-4)J_1^2/\Nsub$ and $\Nsub=4$).~\cite{FN-16tri}
Here, $\Ssub=2$ and we have $2\Ssub+1=5$ degenerate singlet states
(see Appendix~\ref{app-singlet}).

Table~\ref{t-levels} shows the results
For fcc Type-I with $N=32, s=1/2$.
The preceding semiclassical
considerations lead to an energy pattern
which begins with $(1,2)(2,1)$ and ends with $(1,1)(2)$. 
These clusters correspond to orbits $C_0$, $C_1$, $T_1$, and $T_0$.
The total pattern could be written as $(1,2)(2,1/1,1)(2)$, 
where the ``1/1'' indicates a single level which 
can equally well be assigned as
the highest level in cluster $C_1$ or
the lower level of cluster $T_1$.
In Table~\ref{t-levels}, the entries with ``...''
correspond to this level with the ambiguous labeling. 
The diagonalization results (especially for the full
32-site system, see Sec.~\ref{subsec-exactfcc} below)
illustrate how, near the separatrix energy,
the inter- and intra-cluster splittings  become comparable. 

Note that in this paper 
we have not attempted to estimate the tunnel splitting 
(i.e., the splitting within each cluster)
quantitatively, because we have only computed the
``WKB'' exponent in this splitting and not its 
prefactor (all these splittings are exponential
in cluster size).~\cite{4spin,FN-prefactor,FN-WKBerratum}

\subsection{Spatial symmetries}
\label{subsec-symmetries}

The Hamiltonian (\ref{eq-Ham}) is invariant under the
space-group $\cal L$, which acts on the lattice of spins.
Consequently the eigenstates transform as representations
of this symmetry group; in particular, they all
(can be chosen to) have definite crystal momentum 
under lattice translations. 
Similarly, our $p$-spin Hamiltonian (\ref{eq-H2simpler})
is invariant under all permutations of the $p$ spins, 
so its eigenstates transform as the
representations of  the permutation group ${\cal S}_p$.

Now, as observed by Lecheminant {\it et al},~\cite{lech95} 
each space-group operation induces a permutation of the $p$ sublattices. 
Thus the symmetry eigenvalues of the $p$-spin eigenstates
under the permutation group are the same as those of the
corresponding eigenstates of the real Hamiltonian under
the corresponding space operation.  Ref.~\onlinecite{lech95}
combined this idea with group theory to count the number of
times each representation appears among the $p$-spin singlets.
Here, we outline how the same notion may be combined with
semiclassics to identify {\it which}
level has which quantum number,  including the specific
correspondence for both the triangular and fcc realizations of $p=4$. 
That will require an additional mapping from $p$-spin permutations
to (certain) cubic-symmetry point operations on the $\nn$-vector 
in (\ref{eq-H2onespin}). 

Our $p$-spin approximation groups the $\Nsub$ spins of one
sublattice into one big spin of maximum length, i.e.
our states are totally symmetric under any permutation of sites
within a sublattice. 
Thus we implicitly
assumed that the symmetry eigenvalue is unity under 
any translation that takes spins into the same sublattice.
In effect, such translations are equivalenced to the identity, 
and the entire space group reduces to ${\cal S}_p$
(or a subgroup). 



In the four-sublattice case, every lattice translation is equivalent 
(if not to the identity) to a permutation of the class
$(12)(34)$.  
(We write permutations in the standard notation
where each parenthesis is a cyclic permutation and 
each number is the label of a sublattice.)
But for any configuration of the four classical 
directions $\{ \mm_\alpha \}$, 
that permutation can be implemented
instead by a $\pi$-rotation about the axis $\nn_3$. 
Since we defined our coherent state $|\{ \mm_\alpha \}\rangle$
to be symmetrized with respect to all rotations, it follows
that the state must be {\it even} under permutations of the
(12)(34) class.~\cite{FN-kzero}
One corollary is that, in the 4-spin case, the low-lying
singlets {\it all} have wavevector zero, as indeed was
found in the exact diagonalizations of the triangular~\cite{lech95}
and of the fcc type I (Sec.~\ref{subsec-exactfcc}). 

Another corollary is that all permutations of this class are
equivalenced to the identity; in effect, our
permutation group is replaced by ${\cal S}_4/{\cal Q} \cong {\cal S}_3$ 
where ${\cal Q}$ is the subgroup consisting of $(12)(34)$, $(13)(24)$, 
$(14)(23)$ and the identity. 

The permutations of class $(123)$ correspond to the threefold axes in
the triangular lattice or the fcc lattice, as well as about
the threefold axes $(\pm 1,1,1)/\sqrt 3$ of the $\nn$-sphere 
under Hamiltonian (\ref{eq-H2onespin}). 
Group theory suffices to identify 
the behavior of each state under (123): the nondegenerate states have
eigenvalue $1$ and the twofold degenerate states have eigenvalues
$e^{\pm 2\pi i/3}$. 

However, group theory is insufficient to identify the eigenvalues for
the nondegenerate states under the odd permutations, 
which might be $+1$ or $-1$. The permutations
of class $(12)$ correspond to any lattice reflection
in the triangular lattice, or reflections in a 
$\{110\}$-type plane for the fcc lattice.
The permutations of class $(1234)$ correspond to glide
planes in either lattice, but they are equivalent to the $(12)$ class
modulo subgroup $\cal Q$. On the $\nn$-sphere, 
an odd permutation is best represented by
a $\pi/2$-rotations about one of the three coordinate axes. 
(Recall that the $\pi$-rotations are equivalenced to the identity.)

Consider the eigenstates in a three-cluster, mixed from three BS orbits 
$C_{lX}$, $C_{lY}$, and $C_{lZ}$
(these are collinear states rotating 
around the $X$, $Y$, and $Z$ axes respectively).
The wavefunction described by $C_{lX}$ picks up a phase
$4\pi l$ when it goes full circle,~\cite{4spin} thus 
it gets multipied by $e^{i\pi l} = (-1)^l$ under a $\pi/2$
rotation about the $X$ axis (that orbits center). 
The nondegenerate eigenstate has an equal admixture 
of all three states, in particular $C_{lX}$,  so it 
must have eigenvalue $(-1)^l$ under this rotation, or
equivalently under the 
permutation (23).~\cite{FN-inducedrep}
Thus, the
lone eigenstate in the lowest cluster is even
under odd permutations of sublattices, and alternates
odd/even from then on.  The twofold degenerate
eigenstate contains can be broken into one odd
and one even state (the state with eigenvalue $-(-1)^l$
under (23) is a mixture only of $C_{lY}$ and $C_{lZ}$).

Numerical diagonalization of the four-spin system
(for all $\Ssub \leq 13/2$) confirms the predicted pattern
of even/odd states (see Sec.~\ref{subsec-4spindiag}).
In the exact-diagonalization results, it can only be checked 
in the case of the $N=16$ triangular lattice 
(which agrees).~\cite{FN-oddperm}
Empirically, the even/odd  alternation (of the nondegenerate
states) continues across the separatrix to the
higher-energy clusters composed of two states, which 
are mixed from the two BS orbits called $T_{l\pm}$. 
It turns out that whenever such a pair is split, 
the lower level is even (odd) and the upper level
is odd (even) under the odd permutations, according
to whether $l$ is even (odd).

\subsection{Dependence on size $N$}
\label{subsec-N}

How do our results relate to the macroscopic limit $N \to \infty$?
It has long been understood~\cite{and52} that 
in quantum antiferromagnets symmetry-broken states 
(those with a nonzero expectation of each spin)
are superpositions of a family of eigenstates 
which become degenerate with the ground state only
as $N\to \infty$.  This motivates a classification of the
smallest energy gaps, measured from the ground state.

In the present case of a four-sublattice system,  
the smallest gap is within
a cluster of three singlets
(as derived in Ref.~\onlinecite{4spin})
which are mixed from the lowest 
Bohr-Sommerfeld orbits ($C_{X0}, C_{Y0}, C_{Z0}$)
as discussed in Sec.~\ref{subsec-BS}. 
As discussed above, this gap is a tunnel splitting
\cite{FN-qunucl}
and is exponentially small in $N$. 
(The same is true for any discrete symmetry breaking.
But this gap is {\it larger} than in many other cases, since
the tunnel barrier comes entirely from quantum fluctuations.) 

The other gaps are associated in a well-known fashion
with the continuous symmetries or degeneracies
of the classical system. 
If a continuous symmetry is broken and  the interactions are
short-range, there are always gapless ``Goldstone mode''
excitations, in this case the acoustic magnons.
The states with one long-wavelength magnon (at a nonzero 
wavevector $\bf k$) have much larger energies, $E = O(1/N^{1/d})$ 
as $N \to \infty$ in a $d$-dimensional system. 
This follows from the antiferromagnetic magnon dispersion 
$\omega = O(|{\bf k}|)$ and from $|{\bf k}|= O(1/N^{1/d})$
for the smallest wavevector  (with periodic boundary conditions). 
Our present theory does not address the magnon states, and their
gap is not so small anyhow in a finite system. 

The limit of the acoustic magnon as ${\bf k}=0$ has
zero energy, but only in the $N\to\infty$ limit.
It is better to view it from the classical formula 
       $E = M^2/ 2 N \chi$
where $M$ is the total magnetization and 
$\chi$ is  the susceptibility (per spin) in the
macroscopic limit.
Then provided $S_{\rm tot}/N \ll 1$, 
the quantum state of total spin $S_{\rm tot}$ 
has energy $E = S_{\rm tot} (S_{\rm tot} +1)/ 2 N \chi$. 
This ``tower'' of multiplets is generic whenever the system breaks
spin rotation symmetry and the ground state is a 
singlet.~\cite{and52,tower,ber92}
These states have total wavevector 0, but $S_{\rm tot}>0$. 
Their gap is much smaller than the magnons';
it is the next-smallest kind of gap after the tunnel splitting. 
lowest state of spin 1. 

Finally, just as the rotational symmetry breaking implies the
Goldstone modes, 
the additional continuous degeneracy of our four-sublattice
classical ground states implies a gapless ``degeneracy mode''
at harmonic order, with $\omega_{\rm degen}({\bf k}=0) =0$. 
However, quantum fluctuations -- approximated
by our biquadratic effective Hamiltonian (\ref{eq-H2simpler})
-- break that degeneracy,  creating a gap~\cite{bruckel,kim}
$\omega_{\rm degen}({\bf k}=0)= \Delta$ 
this gap becomes constant as $N\to \infty$.

On the other hand, $\Delta$ is exactly 
the energy spacing between successive Bohr-Sommerfeld orbits
surrounding a single collinear state on the $\nn$-sphere
(orbits called ``$C_{lX}$'' in Ref.~\onlinecite{4spin}). 
It can be verified that our semiclassical prescription
indeed implies a constant limiting $\Delta$ 
in the large $N$ limit. 
Expanding the ``one-spin'' effective Hamiltonian 
(\ref{eq-H2onespin}) about a 
minimum point (e.g. the north pole), the $l$-th orbit has
energy $U_l = U_0 + 16 K_U \theta_l^2$, 
where $\theta_l$ is the radius of that (nearly circular) orbit
on the unit sphere (assuming $S=Ns$ is large enough that
$\theta_l \ll 1$). 
Furthermore the orbit area is $\Omega_l = \pi \theta_l^2$, so
from the BS quantization condition (\ref{eq-BSquantC}) 
we obtain $U_l = U_0 + l \Delta$ where
the gap $\Delta = 64 (K_U/S)$. 

At this point we can trace back the $N$ dependence of the
parameters in the successive versions into which the Hamiltonian 
has been mapped (see Table \ref{t-Hams}). 
Writing ``$J$'' to schematically represent all couplings, 
and ``$z$'' to represent the total coordination number, 
we get $\tJ_S=O(Jz/\Nsub)$, 
$K_{12} = O(J^2 z)$, 
$\tilde{K} = O(J/8 \Ssub^3)$, and $K_U = O(J \Ssub/8) = O(J S/32)$. 
Thus finally, $\Delta = O(2 J)$, 
constant to leading order in $1/N$. 

The importance of the above discussion is that
at $N=16$, $28$, or $32$, the apparent $\Delta$ 
is still strongly size-dependent, and is smaller
than the ``tower'' gap to the first eigenstate 
with spin $M=1$. 
Only the analytic analysis reveals that the latter
energy scale will become smaller in a large
enough system. 

\section{Numerical results for fcc Type I}
\label{sec-num}

In this section, we present our numerical results
from exact diagonalization for a lattice 
with various levels of approximation 
based on our effective $p$-spin Hamiltonian (\ref{eq-H2simpler}). 
First, we diagonalize it directly for $p=4$ with parameters
suitable for the fcc Type I system.
Then we will compare the predictions from four-spin 
diagonalization and semiclassics with an 
{\it exact} diagonalization of a 32-site system. 
(The exact diagonalization of a 16-site fcc Type I system
gives five degenerate singlet states which are
what we expect from theory (see Sec.~\ref{subsec-cluster}).)
Finally, we will also give numerical results for $N=16$ and $N=28$
triangular lattices, correcting some minor numerical mistakes
in Ref.~\onlinecite{4spin}.

\subsection{Diagonalization of a four-spin system}
\label{subsec-4spindiag}

We have exactly diagonalized the four-spin Hamiltonian 
(\ref{eq-H2simpler}), as in Ref.~\onlinecite{4spin},
representing a fcc lattice with $N=32, s=1/2$ ($\Nsub=8, \Ssub=4$).
As discussed in Ref.~\onlinecite{smart}, the classical
antiferromagnetic Type I ordering occurs
when $J_1>0, J_2<0$. Here we have used $J_1=1, J_2=-1$.
Our code was previously described in Ref.~\onlinecite{houle99}.
In the diagonalization here, we have used the conservation
of total magnetization in the $z$ direction, $S^z_{\rm tot}$, 
to reduce the size of the Hamiltonian matrix, 
and also the (1234) cyclic permutation symmetry among the four spins.
(See the discussion of odd permutations in Sec.~\ref{subsec-symmetries}.)
We identify the singlet states as those which appear
in the sector with total $z$ spin 0, but not in the
sector with total $z$ spin 1.

The result is presented in Table~\ref{t-levels} 
(and in Table~\ref{t-triangular} for the triangular case).
It should be emphasized that this calculation may be done for systems 
much larger than $N=32$. We could handle $\Ssub = 30$
which corresponds to $N=240$ for a spin $1/2$, $p=4$ system.

\subsection{Comparison to a one-spin Hamiltonian}

As we mentioned above, there
are multiple ways to get a semiclassical spin Hamiltonian
from a given quantum spin Hamiltonian.
Different prescriptions agree to the leading order in spin, 
but the higher order corrections are much harder to obtain.
To estimate the error of our semiclassical Hamiltonians
$\Heffscalar(\{\mm_\alpha\})$ (\ref{eq-H2semi}) 
and $\Hone$ (\ref{eq-H2onespin}) with respect
to the quantum four-spin Hamiltonian $\Heff(\{\SS_\alpha\})$
(\ref{eq-H2simpler}), we construct a one-spin 
{\it quantum} Hamiltonian $\HoneQ$ using
spin coherent state $|\hat{\nn}\rangle=|\theta,\phi\rangle$, 
such that $\langle\theta,\phi|\HoneQ|\theta,\phi\rangle=\Hone$.
(In the notation of Ref.~\onlinecite{shankar}, $\HoneQ$ is the quantum
Hamiltonian for which $\Hone$ is the $\HH_Q$ semiclassical
Hamiltonian.) 
$\HoneQ$ and $\Heff(\{\SS_\alpha\})$ then correspond to the same
semiclassical Hamiltonian $\Hone$, and by comparing
their eigenvalues we can estimate the error
introduced in the semiclassical calculation.

Here we need the expectation value of the quantum operator
$\hat{S}_x^4+\hat{S}_y^4+\hat{S}_z^4$ in coherent states.
We use the following identity (see Appendix~\ref{app-quartic}):
	\begin{eqnarray}
	\nonumber
	&&\langle\theta,\phi|
	\hat{S}_x^4+\hat{S}_y^4+\hat{S}_z^4
	-\frac{3}{5}\left(S^2(S+1)^2-\frac{1}{3}S(S+1)\right)
	|\theta,\phi\rangle\\
	\nonumber
	&=&S(S-\frac{1}{2})(S-1)(S-\frac{3}{2})\\
	\nonumber
	&&\quad\left(
	\cos^4\theta+\sin^4\theta\cos^4\phi+\sin^4\theta\sin^4\phi
	-\frac{3}{5}\right)\\
	&=&\frac{S(S-\frac{1}{2})(S-1)(S-\frac{3}{2})}{S^4}
	\left(S_x^4+S_y^4+S_z^4-\frac{3}{5}S^4\right).
	\label{eq-cs}
	\end{eqnarray}
Here we have emphasized quantum operators with $\hat{S}_\mu$,
and in the last line, $S_\mu$ denotes classical spin variables 
($S_\mu=S n_\mu$, where $n_\mu$ is defined in (\ref{eq-H2onespin})
above). We therefore get a one-spin quantum Hamiltonian from
the semiclassical $\Hone$ (\ref{eq-H2onespin}),
	\begin{equation}
	\HoneQ=-K_Q\sum_{\mu=1}^3 \hat{S}_\mu^4+C_Q
	\label{eq-H2onespinquantum}
	\end{equation}
where
	\begin{equation}
	K_Q=\frac{8K_U}{S(S-1/2)(S-1)(S-3/2)},
	\end{equation}
and
	\begin{equation}
	C_Q=\frac{3}{5}\left(S^2(S+1)^2-\frac{1}{3}S(S+1)\right)K_Q
	-\frac{14}{5}K_U+C_U.
	\end{equation}
Note that here $S=4\Ssub$ is the total spin length.

We have diagonalized this one-spin Hamiltonian 
(\ref{eq-H2onespinquantum}) with $S=16$ 
for the $N=32, s=1/2$ fcc system.
It has the same semiclassical approximation
as our four-spin system, except that 
half the eigenstates of the one-spin system
are disallowed in the four-spin system. 
We emphasize that the single-spin system is not a valid 
model for the extended $N$-spin system in {\it any}
approximation. It is included only to illustrate the
errors in the semiclassical approximations 
mentioned above, since exact diagonalization
gives different results for the one-spin and four-spin 
quantum cases. Tables~\ref{t-levels} and \ref{t-triangular} 
contain exact diagonalization results
for this Hamiltonian (\ref{eq-H2onespinquantum}).

\subsection{Exact diagonalization of the $N=32$ system}
\label{subsec-exactfcc}

The lowest-lying states of a cubic 32-spin cluster with {\em fcc} structure
and periodic boundary conditions were calculated by numerical 
diagonalization. This is the smallest cluster that can accomodate all
three types of antiferromagnetic order predicted by mean field theory
(AF-I, AF-II and AF-III), depending on the values of $J_1$ and
$J_2$. It is a consequence of the periodic boundary conditions
that for a given site, the six different next-nearest neighbor sites
become pairwise identical, such that a spin is only coupled to three 
different next-nearest neighbor spins, each with the strength $2 J_2$. 
Our calculations show that this does not affect the type of order 
adopted by the system.~\cite{LefRis00,RisLef98}

The size of the diagonalization problem was reduced by employing the
following symmetries of the Hamiltonian: total magnetization $S_z$, 
translations,
reflections in 
planes of the $\{ 100 \}$ type, 
and the simultaneous flipping of all spins; and working
in subspaces with particular eigenvalues of these symmetries. The spin-flip
symmetry is only useful for $S_z=0$ and the reflections are only useful
if the translational eigenvalues $e^{i\bf k \cdot r}$ correspond to
wavevectors ${\bf k}=0$ or ${\bf k=}(\pi,0,0)$. 
For $S_z=0$ a subspace dimension of 1213429 was obtained, while for
$S_z=1$, where the spin-flip symmetry was not used, the dimension
was 2259363.

Fortunately, the discussion in Sec.~\ref{subsec-symmetries}, 
shows that the predicted low-lying singlets all have ${\bf k}=0$
and hence are included in the subspaces we checked. 
Note also that reflections of the $\{ 100 \}$ type
map each sublattice into itself, i.e. they correspond to the
identity permutation in ${\cal S}_4$, so all the low-lying
states are predicted to be even under this symmetry, 
which is confirmed by our results. 


The lowest lying eigenstates and eigenenergies were calculated using
the Lanczos algorithm~\cite{Lanczos50,Cullum85} with a random seed. If
an energy level is degenerate, a single run of the algorithm will only
find one eigenvector with this energy. In order to determine the 
degeneracy of the lowest levels we ran the algorithm several times,
each time with a seed that was orthogonal to the low-lying states found
in previous runs, so that only the remaining state space was investigated.
Due to numerical instabilities inherent in the algorithm,~\cite{Cullum85} 
the orthogonalization was also applied after each step
of the Lanczos procedure. The program was tested on a 16-spin {\em fcc}
cluster for which eigenvalues and degeneracies can be found 
analytically,  and gave correct results.
Further details of our method are found in Ref.~\onlinecite{LefRis00}.

The exact diagonalization results for the fcc lattice are in 
Table~\ref{t-levels}. In Table~\ref{t-triangular} we also show
the results for the triangular case which was diagonalized in
Ref.~\onlinecite{lech95}, correcting a factor-of-two mistake
in Table~I of Ref.~\onlinecite{4spin}.
(The Hamiltonian used in Ref.~\onlinecite{lech95}
is twice the Hamiltonian in Ref.~\onlinecite{4spin}
and here.)

\section {Conclusion}
\label{sec-discussion}

In summary, we have studied a class of antiferromagnets
such that their classical ground states have a $p$-sublattice
structure and contain nontrivial degeneracies due to frustration. 
We are interested in the low-lying energy clustering patterns
and have used a $p$-sublattice approach following the 
structure of the classical ground states.
A degenerate family of singlet ground states is introduced,
and an effective Hamiltonian is derived to account for the fluctuations
within sublattices and is used to calculate the low-lying
energies of the antiferromagnet. This effective Hamiltonian
couples the sublattice spins and is of the biquadratic form.
It is written explicitly for a finite-size lattice
and is therefore useful for comparison with exact 
diagonalization results.
We diagonalized a 32-site fcc Type-I system and compared its
low-lying singlets with the analytically obtained 
eigenvalue patterns (from the effective Hamiltonian 
by diagonalization and by semiclassical calculations).
Our main analytical results are summarized in Tables~\ref{t-Heffparams}
and \ref{t-Hams}, and our numerical results are in
Tables~\ref{t-levels} (for fcc) and \ref{t-triangular} (for triangular
lattice). In Table~\ref{t-triangular}, for
the triangular lattice with $N=16$ and $28$, we have corrected
some numerical mistakes in Table I of our earlier paper
Ref.~\onlinecite{4spin}, resulting in a much better agreement
between exact diagonalization and semiclassical results.
As seen in Tables~\ref{t-levels} and \ref{t-triangular}, our method
gives good agreement between theory and numerics for cluster splittings,
but with an overall shift for absolute energy values.

It seems likely that our general approach could be 
applied to catalog the energy and symmetry eigenvalues 
in exact diagonalizations of some other lattice antiferromagnets.
By this approach, we mean (1) representing the low-energy classical states
by several sublattice spins, with lengths proportional to the
size $N$;
(2) writing an effective Hamiltonian for the sublattice spins;
and (3) analyzing this semiclassically, with especial attention
to geometical phases. 
A caution is that the results are likely to be interesting 
mainly when there are multiple tunneling paths, and
when the tunnel barriers  are small compared to other energies
(due either to classical degeneracies broken by quantum fluctuations, 
or else to  weak anisotropies).

The motivation for a careful study of the clustering patterns
of low-lying eigenstates, and thus for this paper, 
is that the pattern might be diagnostic of the ultimate long-range order,
even in systems so small that correlation lengths and
order parameters are inconclusive. 
Size is the key limiting factor to the 
usefulness of exact diagonalization 
for quantum many-body systems, so any approach that partially overcomes
this is of interest. 
The eigenvalue pattern was decisive
evidence of long-range order for the case of 
the $s=1/2$ nearest-neighbor 
triangular antiferromagnet.~\cite{ber92}

Even richer clustering patterns occur in 
systems with classical ground state degeneracies and/or
discrete symmetry breakings, as treated in this paper. 
However, studies~\cite{houle99} subsequent to Ref.~\onlinecite{4spin}
cast this partially into doubt: the eigenvalue pattern
of the four-spin problem with full rotational symmetry reappeared
for three spins with XY symmetry, and presumably 
in many systems with a discrete threefold symmetry
(see Sec. V A of Ref.~\onlinecite{houle99}). 

An alternate application of this theory would be to
the burgeoning topic of small magnetic molecules, 
the magnetic Hamiltonian of which may often be
approximated as a combination of several large subspins.~\cite{ka99}
In the examples studied to date, the intramolecular
interactions were ferromagnetic and highly anisotropic, 
but it is quite plausible that a molecule
(or complex) with high symmetry could realize
a Hamiltonian similar to one of those treated in this paper. 
If furthermore an experimental probe became available to measure a large
number of the energy levels, this subject could be developed
as chemists have developed the semiclassical treatment of
molecular rotational-vibrational levels 
over the past twenty years.~\cite{harter}

It would be desirable to generalize our
derivation of the effective Hamiltonian 
to other systems with classical degeneracies, such as
the fcc Type II or Type III antiferromagnets
(for comparison to the exact diagonalizations of
Ref.~\onlinecite{LefRis00}). 
This does not appear trivial. 
We started by finding the {\it quantum} ground-states of the infinite-range
Hamiltonian $\HMF$, 
and the degeneracy of its ground states 
was essential to our derivation, (the $p$-sublattice singlets) 
particularly in factoring out the energy denominators in 
Subsec.~\ref{subsec-collapse}.

Consider, as the simplest example~\cite{hen89}, 
the square lattice with sufficiently strong antiferromagnetic
$J_2$ that the classical ordering wavevector is $(\pi,0)$ or $(0,\pi)$.
The even spins form two sublattices, oriented oppositely, 
so they combine to make a net {\it singlet}. Likewise the two 
sublattices of odd spins make a singlet, so
$\HMF$ has a unique ground state in this case. (The classical
degeneracy is reflected only in the {\it excited} states 
of $\HMF$.) Rather than pursue such a complicated derivation, 
one could instead just derive the effective 
Hamiltonian as in Ref.~\onlinecite{larson}
(see our Eq.~(\ref{eq-lh})), but this does not
capture the system-size dependence of the coefficients.

\acknowledgments
This work was supported by NSF grants 
DMR-9981744 (at Cornell).
C.L.H. thanks G.~S.~Ezra and B.~Bernu for useful discussions.

\appendix
\section{Adding spins to get singlets}
\label{app-singlet}

To add up $p$ spins of length $\Ssub$ to get a singlet,
we use the well-known decomposition of the product
of two representations: ${\cal D}^{S_1}
\times{\cal D}^{S_2}={\cal D}^{|S_1-S_2|}+{\cal D}^{|S_1-S_2|+1}+...
+{\cal D}^{S_1+S_2}$, where ${\cal D}^S$ is the spin $S$ irreducible
representation of $SU(2)$. (Note that to get a singlet (${\cal D}^{0}$)
in the preceeding expression, we must have $S_1=S_2$.)
Here we show a somewhat nontrivial
calculation for $p=5$ (and assume $\Ssub$ is an integer).
One can either find the coefficent
of ${\cal D}^{\Ssub}$ in ${\cal D}^{\Ssub}\times{\cal D}^{\Ssub}
\times{\cal D}^{\Ssub}\times{\cal D}^{\Ssub}$, or try
$({\cal D}^{\Ssub}\times{\cal D}^{\Ssub}\times{\cal D}^{\Ssub})
\times({\cal D}^{\Ssub}\times{\cal D}^{\Ssub})$.
Adopting the latter approach, and denoting $N_S$ the coefficent
of ${\cal D}^{S}$ in 
${\cal D}^{\Ssub}\times{\cal D}^{\Ssub}\times{\cal D}^{\Ssub}$,
we get $N_S=2S+1$ for $0\le S \le \Ssub$
and $N_S=3\Ssub-S+1$ for $\Ssub+1 \le S \le 3\Ssub$.
The number of singlets for $p=5$ is
thus $\sum_{S=0}^{2\Ssub}N_S=5\Ssub(\Ssub+1)/2+1$. (The upper limit
of the sum goes to $2\Ssub$ because the maximum spin length
from ${\cal D}^{\Ssub}\times{\cal D}^{\Ssub}$, the last two
terms in the product of five terms, is $2\Ssub$.) 

\section{Quartic operator and spin coherent state}
\label{app-quartic}

Here we note briefly how we compute
the spin coherent state expectation value in (\ref{eq-cs}).
The straightforward way is to use a generating function
(Eq.~(6.22) in Ref.~\onlinecite{radcliffe}).
We show here a second way 
using tensor operators (Ref.~\onlinecite{gilmore}, Eq.~(8Q)):
	\begin{equation}
	\langle\theta,\phi|{\cal Y}_{LM}(\SS)|\theta,\phi\rangle=
	\frac{(2S)!}{(2S-L)!2^L}Y_{LM}(\theta,\phi).
	\label{eq-csoperator}
	\end{equation}
Here $Y_{LM}(\theta,\phi)$ is the usual spherical harmonic function.
${\cal Y}_{LM}(\SS)$ is the spherical tensor operator
obtained from the polynomial $S^L Y_{LM}(\theta,\phi)$ by first
substituting
the operators $\hat{S}_x$, $\hat{S}_y$, and $\hat{S}_z$
for classical numbers
$S_x$, $S_y$, and $S_z$, and then symmetrizing the resulting operator, 
e.g., the polynomial $S_x S_y$ becomes the symmetrized
operator $(\hat{S}_x \hat{S}_y+\hat{S}_x \hat{S}_y)/2$.

This identity tells us that if we can write a spin operator
as a linear combination of 
spherical tensor operators ${\cal Y}_{LM}(\SS)$,
then its coherent state expectation value is easy to obtain.
For our problem, it is easy to check the following classical identity:
	\begin{eqnarray}
	\nonumber
	&&S_x^4+S_y^4+S_z^4-\frac{3}{5}S^4
	=\sqrt{\frac{4\pi}{9}}S^4\\
	&&\left[
	\frac{2}{5}Y_{4,0}(\theta,\phi)
	+\frac{2}{\sqrt{70}}\left(
	Y_{4,4}(\theta,\phi)+Y_{4,-4}(\theta,\phi)\right)\right].
	\end{eqnarray}
Then according to (\ref{eq-csoperator}), we have
	\begin{eqnarray}
	\nonumber
	&&\langle\theta,\phi|
	\sqrt{\frac{4\pi}{9}}
	\left[
	\frac{2}{5}{\cal Y}_{4,0}(\SS)
	+\frac{2}{\sqrt{70}}\left(
	{\cal Y}_{4,4}(\SS)+{\cal Y}_{4,-4}(\SS)\right)
	\right]
	|\theta,\phi\rangle\\
	&=&\frac{S(S-\frac{1}{2})(S-1)(S-\frac{3}{2})}{S^4}
	\left(S_x^4+S_y^4+S_z^4-\frac{3}{5}S^4\right).
	\end{eqnarray}
What is the operator on the left hand side of the 
previous equation? Using the table in Ref.~\onlinecite{hutchings},
it can be checked that this spin operator is nothing but
	\begin{equation}
	\hat{S}_x^4+\hat{S}_y^4+\hat{S}_z^4
	-\frac{3}{5}\left(S^2(S+1)^2-\frac{1}{3}S(S+1)\right).
	\end{equation}
On the other hand, we see that the three spherical harmonic
functions involved here all have $L=4$ and according to 
(\ref{eq-csoperator}) the coefficient in front of the
expectation value depends on $L$ only, i.e., the 
coherent state expectation values
of the corresponding spherical tensors have the same coefficients.
We can then start from the classical expectation value
$S_x^4+S_y^4+S_z^4-\frac{3}{5}S^4$ and find the
corresponding operator. It is straightforward to get
$\hat{S}_x^4+\hat{S}_y^4+\hat{S}_z^4$, but because of
operator symmetrization,
$S^4$ goes to $S^2(S+1)^2-S(S+1)/3$ (see Ref.~\onlinecite{hutchings}, p.268,
also Ref.~\onlinecite{yosida}, p.31).
Putting these two terms together, we get the identity
(\ref{eq-cs}) in our paper.

\newpage
\onecolumn

\begin{table}
\caption{
Parameters for the effective Hamiltonian}
\label{t-Heffparams}
\begin{tabular}{ccc}
General & Triangular & Fcc \\
\hline
$p$ & 4 & 4\\
$z^1_{\alpha\beta}$ ($\alpha\ne\beta$) & 2 & 4 \\
$z^2_{\alpha\beta}$ ($\alpha\ne\beta$) & 2 & 0 \\
$z^1_{\alpha\alpha}$ & 0 & 0 \\
$z^2_{\alpha\alpha}$ & 0 & 6 \\
$\tJ_{12}=(\sum_m J_m z^m_{12})/\Nsub$ & $2(J_1+J_2)/\Nsub$ & $4J_1/\Nsub$ \\
$\tJ_{11}=(\sum_m J_m z^m_{11})/\Nsub$ & 0 & $6J_2/\Nsub$ \\
$\EMF=(p/2)(\tJ_{11}\Ssub^2-\tJ_{12}\Ssub(\Ssub+1))$
& $-4(J_1+J_2)\Ssub(\Ssub+1)/\Nsub$ & 
$-8J_1\Ssub(\Ssub+1)/\Nsub+12J_2\Ssub^2/\Nsub$ \\
$K_{\alpha\beta}=\sum_m(J_m)^2 z^m_{\alpha\beta}
-\left(\sum_m J_m z^m_{\alpha\beta}\right)^2/\Nsub$
& $2(J_1^2+J_2^2)-4(J_1+J_2)^2/\Nsub$
& $4(\Nsub-4)J_1^2/\Nsub$
\end{tabular}
\end{table}

\begin{table}
\caption{
The series of Hamiltonians}
\label{t-Hams}
\begin{tabular}{ccccc}
No. of spin & Spin & Hamiltonian & Parameters & Eqn. No. \\
\hline
$N$ & $s$ & $\Hop(\{\ss_i\})=\frac{1}{2}\sum_{ij}J_{ij}\,\ss_i\cdot\ss_j$
& $J_m$ & (\ref{eq-Ham}) \\
& & & $\tJ_S=\tJ_{12}-\Nsub\tJ_{11}/(\Nsub-1)$ & \\

$\Nsub=N/p$ & $\Ssub=\Nsub s$ & 
$\Heff(\{\SS_\alpha\})=-\tilde{K}\sum_{\alpha<\beta}(\Sab)^2+\tilde{C}$ &
$\tilde{K}=K_{12}/(2\tJ_S\Nsub\Ssub(2\Ssub-1)^2)$ &
(\ref{eq-H2simpler}) \\
& & & $\tilde{C}=-p\tilde{K}\Ssub^2((p+1)\Ssub^2-2(p-1)\Ssub-2)/2$ & \\

1 & $S=Ns$ & 
$\Hone=-8K_U\sum_{\mu=1}^3 n_\mu^4+2K_U+C_U$ &
$K_U=\tilde{K}\Ssub^2(\Ssub-1/2)^2$ &
(\ref{eq-H2onespin}) \\
& & & $C_U=-p\tilde{K}\Ssub^2(4\Ssub(p-1)+p+1)/8+\tilde{C}$ & \\

1 & $S$ &
$\HoneQ=-K_Q\sum_{\mu=1}^3 \hat{S}_\mu^4+C_Q$ &
$K_Q=8K_U/(S(S-1/2)(S-1)(S-3/2))$ &
(\ref{eq-H2onespinquantum}) \\
& & & $C_Q=3\left(S^2(S+1)^2-S(S+1)/3\right)K_Q/5-14K_U/5+C_U$ &

\end{tabular}
\end{table}

\begin{table}
\caption{Eigenenergies $\Delta E=E-E_{MF}$ for $N=32$ fcc Type I
($J_1=1,J_2=-1$).
(Entries ``...'' denote a level where cluster labeling ambiguity occurs and
are explained in Sec.~\ref{subsec-cluster}.)
}
\label{t-levels}
\begin{tabular}{lllll}
Orbit, degeneracies  & {\hfil Exact(\ref{eq-Ham}) \hfil}  
& 4-spin  (\ref{eq-H2simpler}) &  1-spin (\ref{eq-H2onespinquantum})
& Semiclassic (\ref{eq-H2onespin}) \\
      & (mean, levels) & (mean, levels) & (mean, levels) & (mean) \\
\tableline
$T_0$ 2 &
-1.5388 &
-1.0316 &
-1.0659 &
-1.0727 \\

$T_1  	\cases{1&\cr 1&\cr}$ & 
	$-1.7457 \cases{-1.7126 &\cr \ldots &\cr}$  &
      	$-1.2029 \cases{-1.1492 &\cr \ldots &\cr}$  &  
      	$-1.1898 \cases{-1.1436 &\cr \ldots &\cr}$  & 
	-1.1718 \\

$C_1  	\cases{1&\cr 2&\cr}$ & 
	$-1.8409 \cases{-1.7788 &\cr -1.8720  &\cr}$  &
      	$-1.2849 \cases{-1.2566 &\cr -1.2990  &\cr}$  &
      	$-1.2519 \cases{-1.2361 &\cr -1.2599  &\cr}$  & 
	-1.2899 \\

$C_0  	\cases{2&\cr 1&\cr}$ & 
	$-2.3178 \cases{-2.3151 &\cr  -2.3233 &\cr}$  &
      	$-1.6706 \cases{-1.6703 &\cr  -1.6713 &\cr}$  &  
      	$-1.5640 \cases{-1.5639 &\cr  -1.5642 &\cr}$  & 
	-1.5639 
\end{tabular}
\end{table}

\begin{table}
\caption{Eigenenergies $\Delta E=E-E_{MF}$ 
for $N=16$, $N=28$ triangular lattices
($J_1=1,J_2=0.7$). 
}
\label{t-triangular}
\begin{tabular}{lllll}
Orbit, degeneracies  & {\hfil Exact(\ref{eq-Ham}) \cite{lech95} \hfil}  
& 4-spin  (\ref{eq-H2simpler}) &  1-spin (\ref{eq-H2onespinquantum})
& Semiclassic (\ref{eq-H2onespin}) \\
      & (mean, levels) & (mean, levels) & (mean, levels) & (mean) \\
\tableline
$\quad N=16$ &&&&\\

$T_0$ 2 &
-0.0756 &
-0.0730 &
-0.0792 &
-0.0824 \\

$C_0  	\cases{2&\cr 1&\cr}$ & 
	$-0.1359 \cases{-0.1331 &\cr -0.1417 &\cr}$  &
      	$-0.1366 \cases{-0.1343 &\cr -0.1412 &\cr}$  &  
      	$-0.1178 \cases{-0.1170 &\cr -0.1193 &\cr}$  & 
	-0.1176 \\
&&&&\\
$\quad N=28$ &&&&\\

$T_0$ 2 &
-1.8769 &
-1.9191 &
-1.9929 &
-2.0135 \\
$C_1  	\cases{2&\cr 1&\cr}$ & 
	$-2.0106 \cases{-2.0022 &\cr -2.0275 &\cr}$  &
      	$-2.3361 \cases{-2.2993 &\cr -2.4097  &\cr}$  &
      	$-2.2821 \cases{-2.2601 &\cr -2.3261  &\cr}$  & 
	-2.3699 \\

$C_0  	\cases{1&\cr 2&\cr}$ & 
	$-2.3843 \cases{-2.3840 &\cr -2.3845 &\cr}$  &
      	$-3.1576 \cases{-3.1539 &\cr -3.1595 &\cr}$  &  
      	$-2.9257 \cases{-2.9246 &\cr -2.9262 &\cr}$  & 
	-2.9252
\end{tabular}
\end{table}



\begin{references}
\bibitem{FN-notkag}
The theory in this paper does not apply to 
cases where the ground state has a macroscopic number of degrees
of freedom, such as the kagom\'e or pyrochlore lattices 
with nearest-neighbor interactions.
\bibitem{FN-fermiliquid}
The Landau Fermi liquid theory, like our theory, focuses on the
energy levels.  The sublattice magnetization renormalization
is like the wavefunction renormalization $Z$ (the overlap of the
bare creation operator with the quasiparticle creation operator):
$Z$ can be small compared to 1, implying a strong renormalization
of the quasiparticles, and yet the dispersion keeps the same
linear dispersion as the noninteracting theory.  One can relate
the noninteracting theory to the Fermi liquid by an adiabatic turn-on
of interactions; in our model, the turn-on of 
$\delta {\cal H}$. One expects that the eigenenergies are proportional 
to the corresponding eigenenergies of the mean-field 
(or respectively non-interacting) systems, but the coefficients
in our effective Hamiltonian (respectively the Fermi liquid $m^*$)
might be strongly changed.
\bibitem{FN-othersphere}
There appears to be an alternate, one-to-one representation 
a single of $\{ \mm_\alpha \}$ by a single sphere, 
as shown in Fig.~6 and footnote 26 of Ref.~\onlinecite{houle99}.
This captures the topology of the phase space and tunneling paths,
but the effective Hamiltonian on this sphere has not been 
related quantitatively to the one derived in the present paper.
\bibitem{FN-patternerror}
The text of Ref.~\onlinecite{4spin} erroneously
reversed the (1,2) and (2,1) pattern, due to a 
sign error in the eigenenergy formula for the $C$ orbits
on p. 5223 of that paper: it should have read 
$U_{Cl}- 2|t_{Cl} \cos[(2\pi j- {\rm Re}\Phi_{Cl})/3]$.
\bibitem{FN-16tri}
For $N=16$ in the triangular case, the
energy scale of the effective Hamiltonian is
anomalously small but nonzero, both in the analytic theory and in 
the exact diagonalization.
\bibitem{FN-WKBerratum}
Ref.~\onlinecite{4spin} contains (at bottom of p.~5223), 
a formula for the WKB exponential
``$(8K_U)\exp (-S {\rm Im} \Phi_{Cl}/3)$.''
This should be corrected by removing the ``$S$'' factor, since 
that was already included in the definition of $\Phi_{Cl}$.
\bibitem{FN-inducedrep}
This argument is inspired by the notion of ``induced representation''
in Ref.~\onlinecite{harter}, that tells us 
which representations of the full symmetry group contain 
a chosen representation of the subgroup 
generated by (23).
\end{references}
\end{document}